\documentclass[twocolumn]{aastex63}
\usepackage{amsmath,xcolor}

\usepackage{soul}
\sethlcolor{green}
\definecolor{mypink}{RGB}{219, 48, 122}

\submitjournal{ApJ}

\shorttitle{Spatially resolved cool galactic winds at $z=1.3$}
\shortauthors{W. Wang et al.}

\begin{document}

\title{The Baltimore Oriole's Nest: Cool Winds from the Inner and Outer Parts of\\ a Star-Forming Galaxy at $z=1.3$}

\author[0000-0002-9593-8274]{Weichen Wang}


\affiliation{Department of Physics and Astronomy, Johns Hopkins University, 3400 N. Charles Street, Baltimore, MD 21218, USA}

\author{Susan A. Kassin}


\affiliation{Department of Physics and Astronomy, Johns Hopkins University, 3400 N. Charles Street, Baltimore, MD 21218, USA}

\affiliation{Space Telescope Science Institute, 3700 San Martin Drive, Baltimore, MD 21218, USA}

\author{S. M. Faber}

\affiliation{University of California Observatories and Department of Astronomy and Astrophysics, University of California, Santa Cruz, 1156 High Street, Santa Cruz, CA 95064, USA}

\author[0000-0003-3385-6799]{David C. Koo}

\affiliation{University of California Observatories and Department of Astronomy and Astrophysics, University of California, Santa Cruz, 1156 High Street, Santa Cruz, CA 95064, USA}

\author[0000-0002-6993-0826]{Emily C. Cunningham}
\affiliation{Center for Computational Astrophysics, Flatiron Institute, 162 5th Ave., New York, NY 10010, USA}

\author{Hassen M. Yesuf}

\affiliation{Kavli Institute for Astronomy and Astrophysics, Peking University, Beijing 100871, China}

\affiliation{Kavli Institute for the Physics and Mathematics of the Universe, The University of Tokyo, Kashiwa, Japan 277-8583}

\author{Guillermo Barro}

\affiliation{Department of Physics, University of the Pacific, 3601 Pacific Ave, Stockton, CA 95211, USA}

\author[0000-0001-8867-4234]{Puragra Guhathakurta}

\affiliation{University of California Observatories and Department of Astronomy and Astrophysics, University of California, Santa Cruz, 1156 High Street, Santa Cruz, CA 95064, USA}

\author{Benjamin J. Weiner}

\affiliation{MMT/Steward Observatory, 933 N. Cherry St., University of Arizona, Tucson, AZ 85721, USA}

\author{Alexander de la Vega}

\affiliation{Department of Physics and Astronomy, Johns Hopkins University, 3400 N. Charles Street, Baltimore, MD 21218, USA}

\author{Yicheng Guo}
\affiliation{Department of Physics and Astronomy, University of Missouri, Columbia, MO 65211, USA}

\author{Timothy M. Heckman}

\affiliation{Department of Physics and Astronomy, Johns Hopkins University, 3400 N. Charles Street, Baltimore, MD 21218, USA}

\author{Camilla Pacifici}

\affiliation{Space Telescope Science Institute, 3700 San Martin Drive, Baltimore, MD 21218, USA}

\author[0000-0001-9269-5046]{Bingjie Wang}

\affiliation{Department of Physics and Astronomy, Johns Hopkins University, 3400 N. Charles Street, Baltimore, MD 21218, USA}

\author{Charlotte Welker}

\affiliation{Department of Physics and Astronomy, Johns Hopkins University, 3400 N. Charles Street, Baltimore, MD 21218, USA}

\correspondingauthor{Weichen Wang}
\email{wcwang@jhu.edu}

\begin{abstract}

Strong galactic winds are ubiquitous at $z\gtrsim 1$. However, it is not well known where inside galaxies these winds are launched from.  We study the cool winds ($\sim 10^4$\,K) in two spatial regions of a massive galaxy at $z=1.3$, which we nickname the ``Baltimore Oriole's Nest.'' The galaxy has a stellar mass of $10^{10.3\pm 0.3} M_\sun$, is located on the star-forming main sequence, and has a morphology indicative of a recent merger. Gas kinematics indicate a dynamically complex system with velocity gradients ranging from 0 to 60 $\mathrm{km}\cdot\mathrm{s}^{-1}$. The two regions studied are: a dust-reddened center (Central region), and a blue arc at 7 kpc from the center (Arc region).   We measure the \ion{Fe}{2} and \ion{Mg}{2} absorption line profiles from deep Keck/DEIMOS spectra. Blueshifted wings up to 450 km$\cdot$s$^{-1}$ are found for both regions. The \ion{Fe}{2} column densities of winds are $10^{14.7\pm 0.2}\,\mathrm{cm}^{-2}$ and  $10^{14.6\pm 0.2}\,\mathrm{cm}^{-2}$ toward the Central and Arc regions, respectively.  Our measurements suggest that the winds are most likely launched from both regions. The winds may be driven by the spatially extended star formation, the surface density of which is around 0.2 $M_\sun\,\mathrm{yr}^{-1}\cdot \mathrm{kpc}^{-2}$ in both regions. The mass outflow rates are estimated to be  $4\,M_\sun\,\mathrm{yr}^{-1}$ and  $3\,M_\sun\,\mathrm{yr}^{-1}$ for the Central and Arc regions, with uncertainties of one order-of-magnitude or more. Findings of this work and a few previous studies suggest that the cool galactic winds at $z\gtrsim 1$ might be commonly launched from the entire spatial extents of their host galaxies due to extended galaxy star formation. 

\end{abstract}



\section{Introduction} \label{sec:intro}

Star-forming galaxies at $z\gtrsim 1$ host strong galactic winds (e.g., \citealt{Lowenthal1997,Frye2002,Adelberger2003, Shapley2003, Tremonti2007, Sato2009, Weiner2009, Rubin2010,Rubin2014, Steidel2010, Bordoloi2011, Bordoloi2014, Coil2011, Erb2012, Kornei2012, Martin2012, ForsterSchreiber2019}). The winds are more ubiquitous at this epoch (e.g., \citealt{Weiner2009}) than in the local universe (e.g., \citealt{Veilleux2005a,Chen2010}). They are known to play an important role in galaxy formation by removing gas and metal from galaxies permanently or temporarily and causing inefficient star formation (e.g., \citealt{Somerville2015}). For the case of temporary removal, winds will be part of the galaxy-halo fountains which allow the ejected gas to eventually return to galaxies (\citealt{Veilleux2005a, Veilleux2020,Somerville2015,Naab2017,Tumlinson2017}).

Winds are one of the essential ingredients hydrodynamic simulations need to incorporate, in order to produce galaxies that match the observed ones in terms of the stellar mass, size, kinematics, and metallicity (e.g., \citealt{Governato2007, Oppenheimer2009, Brook2011,Hopkins2012,Somerville2015,Pillepich2018}). In the simulations, winds are assumed to be launched from individual regions of galaxies that reach a given threshold in star-formation rate density (e.g., \citealt{Oppenheimer2006, Hopkins2012,Vogelsberger2013, Vogelsberger2014, Muratov2015,Grand2019, Nelson2019b, Pandya2021}). Galaxies generated by the simulations, in which these assumptions are applied to individual spatial regions, are found to match the galaxies in the real Universe in the ballpark, not only in terms of the integrated galaxy properties mentioned above but also in terms of some spatially resolved properties (e.g., \citealt{Gibson2013, Belfiore2019, Rodriguez-Gomez2019, Ubler2020, Nelson2021, Simons2021}). Notwithstanding the success, the assumptions about the launching of galactic winds in the simulations need to be directly tested by observations. A first step to perform such a test is to measure whether there are winds launched from individual regions of galaxies using observational data, and (if so) compare the wind properties with the star formation properties of these regions. This needs to be done using deep and spatially resolved observations of galactic winds at a certain cosmic epoch.

Spatially resolved observations of galactic winds are relatively rare at $z\gtrsim 1$, the cosmic epoch when winds are ubiquitous and possibly also their impacts on galaxy formation.  Most previous studies use integrated spectra, due to limitations in spectral sensitivity and spatial resolution (e.g., \citealt{Weiner2009,Rubin2010,Rubin2014,Kornei2012}). There are only about a dozen resolved studies of galactic winds at $z\gtrsim 1$. Most of them focus on the dense and warm ionized phase traced by nebular emission lines. These studies find that the warm ionized winds extend from the galaxy centers to at least 1 $R_e$ (2 to 6 kpc), which indicates that the ionized gas might be launched from an area of several kpc in galaxies (\citealt{Genzel2011, Newman2012, Newman2012b, ForsterSchreiber2014, ForsterSchreiber2019, Davies2019}). 

However, galactic winds exist in multiple phases, including the cool neutral phase (around 10$^{4}$ K), the molecular phase (e.g., \citealt{Veilleux2005a, Wood2015, Baron2020, Fluetsch2018, Fluetsch2020, Roberts-Borsani2020b}), and the ionized phase (e.g., \citealt{Heckman2015,Heckman2017}). 
The gas in the neutral and molecular phases would directly fuel star formation if it were not carried away from galaxies as winds. As a result, winds in the two phases are expected to have the most direct impacts on the galaxy star formation (\citealt{Rupke2018,Veilleux2020}).

To study the cool phase, rest-frame ultraviolet (UV) lines from ions with ionization potentials close to that of the hydrogen atom are needed. At $z\gtrsim 1$, there are no more than a handful of spatially resolved studies using the absorption lines of these ions, including \ion{Mg}{2}, \ion{Fe}{2}, and \ion{Si}{2}. For example, \cite{Bordoloi2016} study a gravitationally lensed galaxy at $z=1.7$ and measured winds toward four bright star-forming clumps of the galaxy. They find that the wind column densities are comparable among the four clumps whereas the wind velocities correlate with the star-formation rate densities of the four clumps.  Two other studies by \cite{James2018} and \cite{RickardsVaught2019} measure the line equivalent widths along multiple sightlines toward a galaxy at $z=2.4$ and a galaxy at $z=0.7$, respectively, and find that the equivalent widths measured from different sightlines are comparable. The galaxy studied by \cite{James2018} is gravitationally lensed whereas the one by \cite{RickardsVaught2019} is not. 
 
In this paper, we build upon the pioneering UV absorption line works cited above and study where cool winds are launched from in an un-lensed star-forming galaxy at $z\sim 1$. We choose to study an un-lensed galaxy to get around the systematics of gravitational lensing: for lensed systems, the star-formation rates (SFRs) are difficult to measure because the magnification factors are subject to substantial uncertainties from lensing models, and typically only a few bright clumps of a target galaxy are detected. 

The galaxy we study comes from a spectroscopic survey on the Keck telescope (see the following section and \citealt{2019ApJ...876..124C,2019ApJ...879..120C}). It has a stellar mass of $10^{10.3}\,M_\sun$ and a typical SFR for its mass, 62  M$_\sun\cdot\mathrm{yr}^{-1}$, and is therefore on the star-forming main sequence (e.g., \citealt{Noeske2007}). It is selected because its spectrum is spatially extended and has a high signal-to-noise ratio in the continuum. We measure wind properties,  including velocities and column densities, for the inner and outer regions of the galaxy from the \ion{Fe}{2} and \ion{Mg}{2} absorption lines, and infer from which region(s) the winds are launched.  

\begin{figure*}
	\centering
	\includegraphics[width = 6. in]{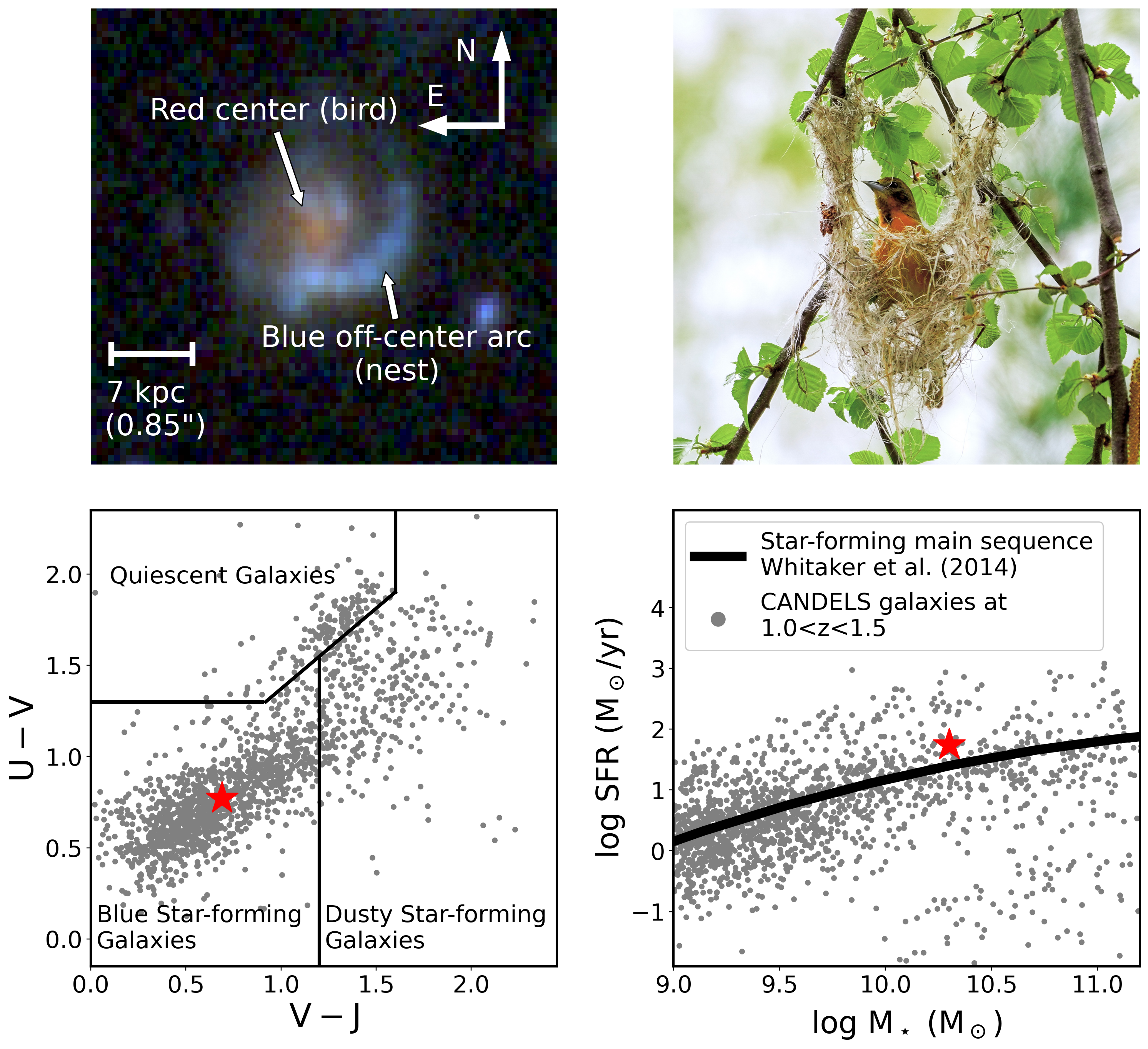}
	\caption{ The Baltimore Oriole's Nest galaxy is a massive star-forming galaxy at $z\sim 1.3$ with highly irregular morphology. \emph{Upper panels:} On the left, the \emph{HST} RGB image (WFC3/F160W for red, ACS/F850LP for green and ACS/F606W for blue) shows a red ``Central" region and a blue off-center ``Arc" that extends from northwest to south.  The galaxy image is analogous to a Baltimore Oriole in its nest, as shown on the right (photo credit: John Anes, 2018).   \emph{Lower panels:} This galaxy is blue and star-forming according to its location on the rest-frame U-V versus V-J diagram, which is marked by a star symbol  (left). The U-V versus V-J diagram is separated into three parts, blue star-forming galaxies, dusty star-forming galaxies and quiescent galaxies, following \cite{Spitler2014}. The galaxy also lies on the star-forming main sequence at $1.0<z<1.5$ (\citealt{Whitaker2014}), as indicated by the red star (right).  Gray points in these lower panels represent individual galaxies at $1.0<z<1.5$ from the GOODS-North field of CANDELS survey.  The rest-frame colors of galaxies are from \cite{Barro2019}. Stellar masses and SFRs are obtained from SED fitting by \cite{Pacifici2012,Pacifici2016}.  \label{fig:sfms}}
\end{figure*}

The paper is structured as follows. Section \ref{sec:targetselection} describes the sample selection, and \S \ref{sec:data} summarizes the spectroscopic observations, reductions, and ancillary data. Section \ref{sec:morphology} discusses the morphology of the galaxy and presents its gas kinematics. Section \ref{sec:extractionspec} describes how we co-add spectra. Section \ref{sec:measuremasssfragn} describes measurements of the SFRs and SFR densities of individual regions of the galaxy.  Section \ref{sec:measurewindline} presents main results of the paper, i.e., measurements of the column densities of winds from inner and outer regions of the galaxy. Section \ref{sec:missingemissionline} discusses possible reasons for the non-detection of the \ion{Fe}{2} and \ion{Mg}{2} emission lines. Based on the results from \S \ref{sec:measurewindline}, \S \ref{sec:discusswindorigin} discusses where inside the galaxy winds are launched from. Section \ref{sec:discussphysicalimpacts} discusses the relation between winds and star formation. The mass outflow rates and mass loading factors are estimated in \S \ref{sec:massoutflowratemassloading}.  A comparison between the results of this study and those of other relevant ones is presented and the prospect of future similar observational studies with the James Webb Space Telescope (\emph{JWST}) is discussed in \S \ref{sec: compare_with_literature}. Conclusions of the paper are given in \S \ref{sec:conclusion}.  Throughout the paper, the wavelengths of spectral lines are from measurements in the air. Quoted magnitudes are in the AB system. A flat $\Lambda$CDM cosmology with $\Omega_M = 0.3$, $\Omega_\Lambda = 0.7$, and a Hubble constant of $H=70\ \mathrm{km}\cdot \mathrm{s}^{-1}\,\mathrm{Mpc}^{-1}$ is adopted.

\section{Selection and Properties of the Baltimore Oriole's Nest galaxy}
\label{sec:targetselection}

\begin{table*}
	\renewcommand{\thetable}{\arabic{table}}
	\caption{Selected Physical Properties of the Baltimore Oriole's Nest galaxy }
	\centering
	\begin{tabular}{llll}
		\hline
		\hline 
		Property        &  Value           & Unit & Reference\\ \hline 
		CANDELS ID & GDN 21734  &  &\cite{Barro2019} \\
		RA (J2000) & 12:37:55.10  &  hh:mm:ss  & \cite{Barro2019} \\
		DEC (J2000) &  62:17:17.88  & dd:mm:ss   &  \cite{Barro2019} \\
		\bf{Global properties} & & & \\
		Spectroscopic redshift &  1.3063   &  & \cite{Barro2019} \\
		\emph{HST}/ACS F606W magnitude &   $22.8\pm0.1$ & AB mag  & \cite{Barro2019} \\
		Effective radius at rest-frame V-band &    $0.50\pm0.05$ ($4.2\pm0.4$) &   arcsec (kpc) & \cite{VanderWel2012} \\
		 (observed \emph{HST}/WFC3 F160W band) & & & \\
		Axis ratio at rest-frame V-band (b/a) & $0.93\pm0.08$  &  & \cite{VanderWel2012} \\
		Stellar mass &   $10^{10.3\pm0.3}$  & $M_\sun$  & \cite{Pacifici2012, Pacifici2016} \\
		Star-formation rate & $62^{+5}_{-31}$ &  $M_\sun\,\mathrm{yr}^{-1}$ & \cite{Pacifici2012, Pacifici2016} \\
		\bf{Resolved properties} & & &  \\
		Redshift (Central region) & 1.3060 & & Appendix \ref{appendix:oiimeasurement} \\
		Redshift (Arc region) & 1.3063 & & Appendix \ref{appendix:oiimeasurement} \\
		Star-formation rate (Central region)  &   $18^{+19}_{-8}$ & $M_\sun\,\mathrm{yr}^{-1}$ & \S \ref{sec:measuremasssfragn} \\   
		Star-formation rate  (Arc region)
		&  $13^{+4}_{-4}$ &   $M_\sun\,\mathrm{yr}^{-1}$ & \S \ref{sec:measuremasssfragn} \\
		Surface area (Central region) & $76$ & $\mathrm{kpc}^2$ & \S \ref{sec:measuremasssfragn}\\
		Surface area (Arc region) & $86$  & $\mathrm{kpc}^2$ & \S \ref{sec:measuremasssfragn} \\
		Star-formation rate density (Central region) &  $0.23^{+0.26}_{-0.10}$ & $M_\sun\,\mathrm{yr^{-1}kpc^{-2}}$       &     \S \ref{sec:measuremasssfragn}  \\
		Star-formation rate density (Arc region) &    $0.15^{+0.04}_{-0.04}$ &  $M_\sun\,\mathrm{\,yr^{-1}kpc^{-2}}$   &  \S \ref{sec:measuremasssfragn} \\
		\hline
	\end{tabular}\label{tab:targetprop}
\end{table*}

The galaxy studied in this work comes from the ``Halo Assembly in Lambda-CDM: Observations in 7 Dimensions'' survey (HALO7D; PI: R.~Guhathakurta; \citealt{Yesuf2017,2019ApJ...876..124C,2019ApJ...879..120C,Pharo2022}), which makes use of the multiplex capability of the DEep Imaging Multi-Object Spectrograph (DEIMOS; \citealt{Faber2003}) on the Keck II Telescope.  HALO7D targets include both Milky Way halo stars and distant galaxies.   All the HALO7D galaxies are within or close to the deep extragalactic fields observed by the Cosmic Assembly Near-infrared Deep Extragalactic Legacy Survey CANDELS (PIs: S.~Faber and H.~Ferguson;  \citealt{Grogin2011,Koekemoer2011}), which provides multi-band imaging data from the Hubble Space Telescope (\emph{HST}).

We select the galaxy because it has a spectrum with a high signal-to-noise ratio in the continuum (S/N = 7 at around rest-frame 2800 \AA) and is spatially extended so that we can measure its wind properties in two spatially distinct regions. It is selected from a sub-sample of star-forming galaxies in HALO7D.  The sub-sample includes galaxies that have spectra which are spatially extended in the continua, larger than 2\arcsec\ in diameter, and have absorption lines that trace the cool galactic winds (\ion{Fe}{2}\,$\lambda\lambda$\,2586/2600\,\AA\ and \ion{Mg}{2}\,$\lambda\lambda$\,2796/2803\,\AA). The galaxy we choose has the highest signal-to-noise ratio in the spectral continuum among the sub-sample. The rest of the sub-sample will be studied in a forthcoming paper.

The \emph{HST} RGB image of the galaxy is shown in upper left of Figure \ref{fig:sfms}, where the following wavebands are used: F160W of the Wide Field Camera 3 (WFC3) for the red channel, Advanced Camera for Surveys (ACS) F850LP for the green channel, and ACS/F606W for the blue channel. We nickname it the Baltimore Oriole's Nest galaxy because it looks similar to a Baltimore Oriole in a nest, which we show in the top right panel of Figure \ref{fig:sfms}\footnote{This bird photo is shared under the CC BY-SA 2.0 License. The original photo by John Anes can be found at \url{https://flic.kr/p/262X4Us}}.  The galaxy's morphology is discussed further in \S \ref{sec:morphology}.

The Baltimore Oriole's Nest galaxy is a blue star-forming galaxy according to its location on the rest-frame U-V versus V-J diagram (\citealt{Spitler2014}), as shown in the lower left panel of of Figure \ref{fig:sfms}.   It is also on the star-formation main sequence at $z\sim1$ \citep{Whitaker2014}, as shown in the lower right panel
of Figure \ref{fig:sfms}.  In these figures, the galaxy is shown as a red star, and the galaxies at $1.0<z<1.5$ in the GOODS-North field of CANDELS are shown as gray points. 
The rest-frame colors of the galaxies are inferred by \cite{Barro2019} using the {\sc eazy} code (\citealt{Brammer2008}).
The integrated stellar masses and SFRs of all the galaxies in the figure are derived from the spectral energy distribution (SED) fitting by \cite{Pacifici2012, Pacifici2016}.

Finally, no active galactic nucleus (AGN) is found in the galaxy, either according to the X-ray criteria by \cite{Xue2011,Xue2016} for unobscured AGNs or the infrared color criteria by \cite{Donley2012} for obscured AGNs.  

A full list of the properties of the Baltimore Oriole's Nest galaxy can be found in Table \ref{tab:targetprop}.

\section{Data \& Data Reduction}

\label{sec:data}

\begin{table*}
	\renewcommand{\thetable}{\arabic{table}}
	\caption{Keck/DEIMOS observations} \label{tab:obslog}
	\centering
	\begin{tabular}{llllll}
		\hline
		\hline
		Observation Date & Slit Position Angle\tablenotemark{a} & Exposure Time & Mask ID & Airmass & Seeing\tablenotemark{b} \\
		& degree & hour &  & & arcsec \\		
		\hline
		2016 Mar 03 
		                        & 10   &   0.33 & GN0b & 1.36 & 0.84 \\
		                        & 10   &   0.33  & GN0b & 1.35 & 0.80 \\
		                        & 10   &   0.30  & GN0b & 1.35 & 0.83 \\
		                        & 10   &   0.30  & GN0b & 1.36 & 0.79 \\
		2016 Mar 03 
		                        & 30  &   0.33  & GN0a  & 1.47 & 0.86 \\
		                        & 30  &   0.33  & GN0a  & 1.43 & 0.85 \\
		                        & 30  &   0.33  & GN0a  & 1.40 & 0.86 \\
		2016 Mar 04 & 10   &   0.33  &  GN0b & 1.37 & 0.78 \\
		                        & 10   &   0.33  &  GN0b & 1.36 & 0.68 \\
		                        & 10   &   0.33  &  GN0b & 1.35 & 0.65 \\
		                        & 10  &   0.33  &  GN0b & 1.35 & 0.67 \\
		2016 Mar 04 & 30  &   0.33  &  GN0a & 1.51 & 0.75 \\
		                        & 30  &   0.33  &  GN0a & 1.39 & 0.85 \\
		2016 Mar 04 & 58  &    0.33 &  GN0c & 1.95 & 0.83 \\
		                        & 58  &    0.33 &  GN0c & 1.82 & 0.78 \\
		                        & 58  &    0.33 &  GN0c & 1.72 & 0.78 \\
		                        & 58  &    0.33 &  GN0c & 1.64 & 0.74 \\
		                        & 58  &    0.25 &  GN0c & 1.57 & 0.64 \\
		\hline
		\multicolumn{4}{c}{Total exposure time: 5.85 hours}\\
		\hline
	\end{tabular}
\begin{flushleft}

\tablenotetext{a}{\noindent The slit position angle is defined relative to north, and it increases from north to east. } 
\tablenotetext{b}{Seeing is measured in the observed V-band from unsaturated alignment stars on each DEIMOS slit mask. Only exposures with seeing $	\leqslant 0.86\arcsec$ are used and listed here.}
\end{flushleft}
\end{table*}

\subsection{Keck/DEIMOS Spectra}

Keck/DEIMOS spectra of the Baltimore Oriole's Nest galaxy were taken during two nights in March, 2016. The 600 line/mm grating centered at 7200\,\AA\ was used along with the GG455 order-blocking filter.  The resulting wavelength range of the spectra is 4600\,\AA\ to 9500\,\AA.  To limit flux losses at shorter wavelengths of the spectra, slit position angles were chosen so that they were no more than $30 \degr$ away from the parallactic angle for each observing session. As a result, the galaxy was observed with slits at three different position angles: 10$\degr$,  30$\degr$ and 58$\degr$, measured from north to east. All slits are 1\,\arcsec\ wide along the dispersion direction, resulting in a spectral resolution of 3.5\,\AA\ full width at half maximum (FWHM).  The resolution $R=\lambda/\Delta \lambda$ is approximately 2100, and the instrumental velocity dispersion is about $70\,\mathrm{km}\cdot \mathrm{s}^{-1}$ \citep{Wirth2004}. 

Seeing varied from 0.7\arcsec\ to 1.0\arcsec\ (FWHM) in the V-band. Only the spectra observed when the seeing was better than 0.86\arcsec\ are used. The total exposure time for these spectra is 5.85 hours. 
A total of 18 individual spectra are used and combined in our analysis.  Their dates of observation, slit position angles, exposure times, airmasses, and the seeing measured as they were observed are given in Table \ref{tab:obslog}. In addition to the galaxy spectra, flat fields were taken with a quartz lamp and wavelength calibration spectra were taken with a blue CdHgZn lamp and a red NeKrArXe lamp.

Spectral reduction is performed using the IDL-based {\sc spec2d} pipeline (\citealt{Cooper2012,Newman2013}).  The pipeline models the sky background as a function of wavelength and subtracts it from the galaxy spectra. The reduction yields a 2D spectrum for each exposure performed at a certain slit position angle. Individual 2D spectra obtained at the same slit position angle are then combined to create a stacked 2D spectrum. 

After stacking, we noticed an additional background contaminant in the spectra that varies with spatial location along the slit, which is not taken into account by the pipeline.  However, this contaminant only contributes no more than 10\% to the spectral flux density relative to the continuum, which is not substantial enough to impact the shapes of the extracted absorption line profiles.  The contaminant is modeled and subtracted nevertheless, for which the detailed steps can be found in Appendix \ref{appendix:backgroundissue}.

 \begin{figure*}[t]
	\flushleft
	\includegraphics[width = 7. in]{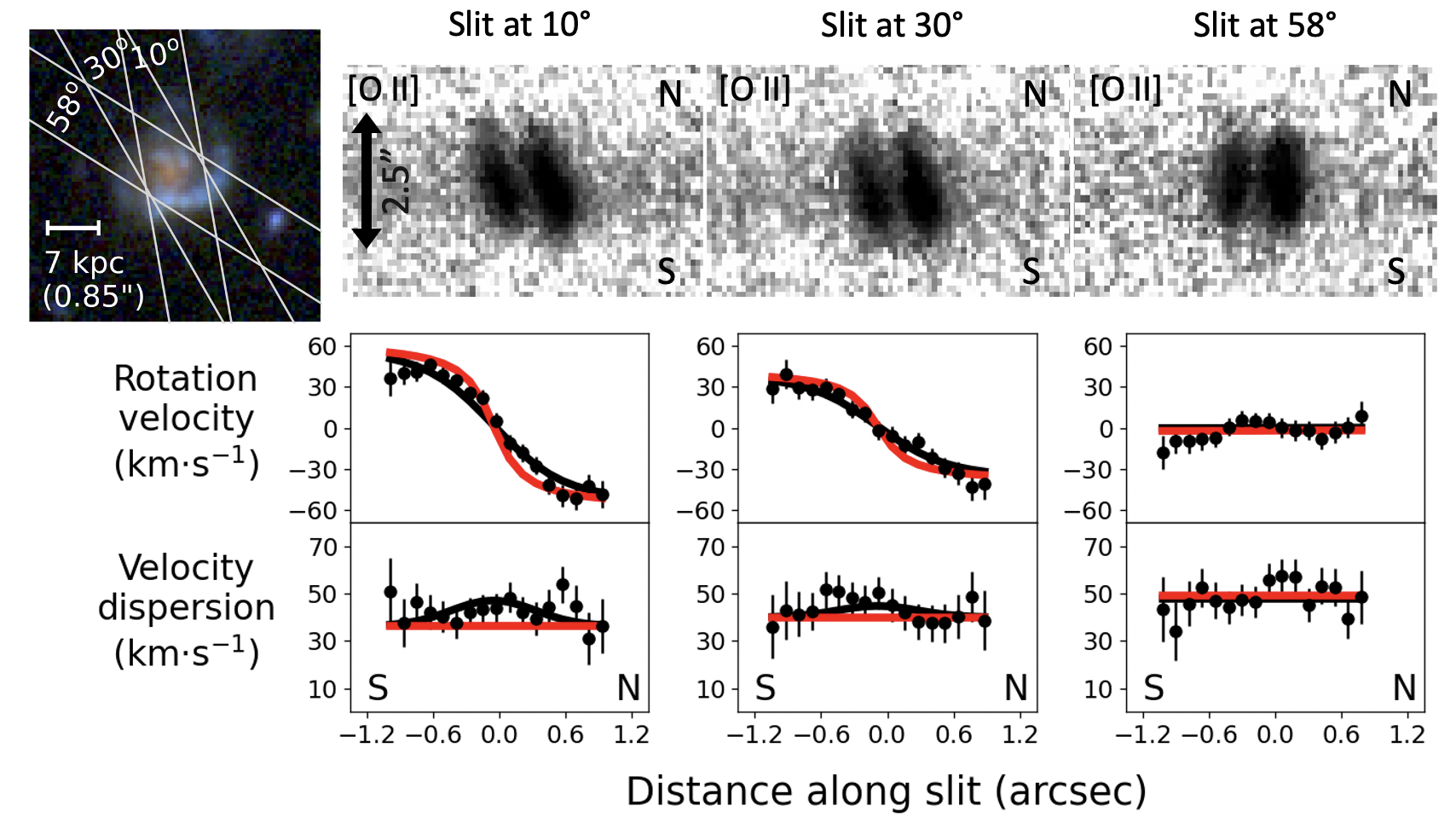}
	\caption{Measurements of gas kinematics for the Baltimore Oriole's Nest Galaxy at 3 slit position angles are shown.  At the upper left is the \emph{Hubble} image of the galaxy with drawings of the slits used to obtain the spectra.  The three black and white images across the top are cut-outs of the [\ion{O}{2}] emission line doublet from the rectified 2D spectra, where the wavelength direction is horizontal and the spatial direction is vertical. We measure gas kinematics from the [\ion{O}{2}] lines following \cite{Weiner2006}. 
	The plots below show the rotation velocity and velocity dispersion as a function of radius, measured from the 2D spectra. The solid points show the results of Gaussian fits to each row of the spectra.  The black lines are the best-fit models to these points, and the red lines are the intrinsic models without the effect of seeing. For the three slit position angles, $10\degr$, $30\degr$, and $58\degr$, the best-fit rotation velocities on the flat part of the rotation curve are $61\pm 4\,\mathrm{km}\cdot\mathrm{s}^{-1}$, $41\pm 3\,\mathrm{km}\cdot\mathrm{s}^{-1}$, and $0\pm 4\,\mathrm{km}\cdot\mathrm{s}^{-1}$, respectively.  A positive velocity indicates that the Arc region is redshifted with respect to the Central region. The intrinsic velocity dispersions, which are assumed to be constant across the face of the galaxy for each position angle, are $36\,\mathrm{km}\cdot\mathrm{s}^{-1}$, $40\,\mathrm{km}\cdot\mathrm{s}^{-1}$, and $48\,\mathrm{km}\cdot\mathrm{s}^{-1}$, respectively, with an uncertainty of around 30 $\mathrm{km}\cdot\mathrm{s}^{-1}$ for each. \label{fig:kinems} }
\end{figure*}

\subsection{Hubble Optical and Near-IR Images}

\label{sec:HSTphotometry}

 \emph{HST} images of the galaxy in multiple wavebands are available from CANDELS. The galaxy is observed in the following bands: ACS F435W, F606W, F775W, F814W, F850LP \citep{Giavalisco2004, Riess2007}, and WFC3 F105W, F125W, F160W \citep{Grogin2011,Koekemoer2011}.  Mosaics for all these wavebands, which are publicly available from the CANDELS data release\footnote{refer to \url{https://archive.stsci.edu/prepds/candels/} and	\dataset[10.17909/T94S3X]{https://doi.org/10.17909/T94}}, are generated using the {\sc MosaicDrizzle} pipeline \citep{Guo2013}. Integrated photometry of each band is performed by running the \textsc{ S\lowercase{extractor}} code \citep{Bertin1996} on the CANDELS images convolved to the resolution of the F160W band, which is 0.18\arcsec\ in FWHM. Details about the photometry are described in \cite{Barro2019}. The integrated fluxes measured for the galaxy are $2.08\pm0.08\,\mu$Jy (F435W),  $2.72\pm0.03\,\mu$Jy (F606W), $4.35\pm0.08\,\mu$Jy (F814W), $6.21\pm0.04\,\mu$Jy (F850LP), $7.64\pm0.11\,\mu$Jy (F105W), $8.62\pm0.07\,\mu$Jy (F125W), and $11.49\pm0.09\,\mu$Jy (F160W). 

\section{Size, Morphology, the Central and Arc regions, \& Kinematics}

\label{sec:morphology}

The galaxy has an effective radius of 4.2 kpc and an axis ratio of 0.93. These quantities are measured by \cite{VanderWel2012} from the \emph{HST}/WFC3 F160W band, which corresponds to the rest-frame V-band, using the GALFIT code \citep{Peng2002}.

\begin{figure*}
	\flushleft
	\includegraphics[width = 7. in]{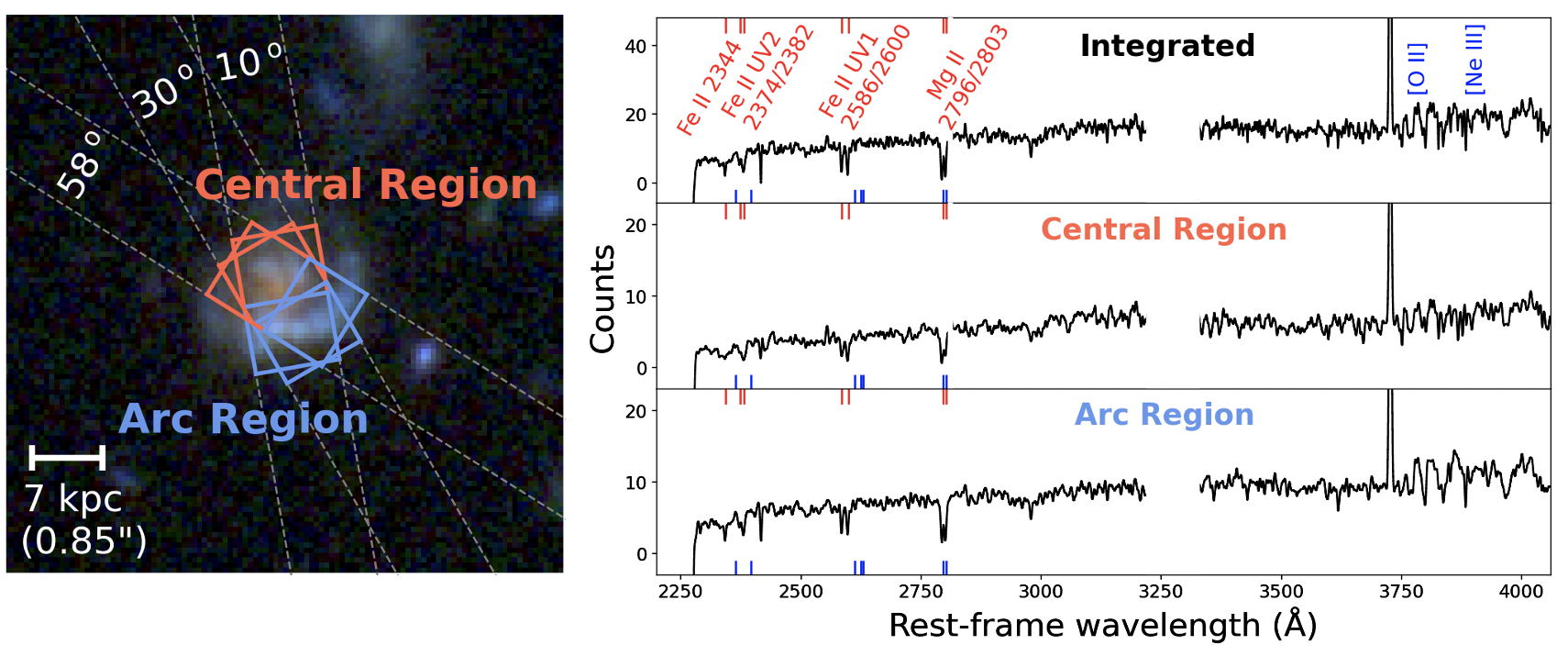}
	\caption{Two spatial regions of the Baltimore Oriole's Nest galaxy are studied in this work, the Central region and the Arc region. Strong \ion{Fe}{2} and \ion{Mg}{2} absorption lines are found for both regions. \emph{Left:} The Central and Arc regions are shown as red and blue rectangles on the Hubble image. The Central region covers the red core of the galaxy and the Arc region covers the blue off-center arc. Spectra observed with three slits (white dashed lines) are extracted and co-added for each region. \emph{Right:}  The integrated co-added spectrum and the co-added spectra extracted from the Central and Arc regions are shown from top to bottom. \ion{Fe}{2} and \ion{Mg}{2} absorption lines, indicated by red ticks, are present in both spatial regions. Blue ticks indicate the wavelengths of \ion{Fe}{2}* and \ion{Mg}{2} emission lines. These lines are absent from the two regions.  The spectra have been smoothed by a Gaussian with an RMS width of 6 DEIMOS pixels, or 1.7 \AA\ in the rest-frame. Wavelengths from 3200\AA\ to 3300\AA\ are impacted by the strong telluric absorption and are removed from the spectra. \label{fig:imageandregion}}
\end{figure*}

The galaxy has a highly irregular morphology.  As the \emph{HST} image in Figure \ref{fig:sfms} shows, it contains a red center, which we refer to as the ``Central region", and a blue extended arc which we refer to as the ``Arc region."  These regions will be precisely defined in the following section.  The Arc region appears to be a tidally distorted structure or a highly disturbed disk, either of which could be caused by a recent merger.  Note that the term ``arc" is used to refer to a morphological feature, and should not be confused with the arc-like images commonly seen in gravitationally lensed systems since the galaxy is not lensed.

The galaxy is categorized as an ``irregular disk galaxy" according to a Deep Learning based morphology classification by \cite{Huertas-Company2015,Huertas-Company2016}.  Specifically, its visual morphology frequency values are $f_\mathrm{irr}=0.70$  and $f_\mathrm{disk}=0.73$ (\citealt{Huertas-Company2015}), which means that there is a 70\% probability that human classifiers would identify the galaxy as ``irregular," and a 73\% probability that they would identify it as ``disky."  The irregularity of the Baltimore Oriole's Nest galaxy, as defined by $f_\mathrm{irr}$, is higher than 80\% of the galaxies with similar values of redshift, stellar mass, and SFR.

Gas kinematics are measured at three separate slit position angles from [\ion{O}{2}] emission lines, as shown in Figure~\ref{fig:kinems}. The footprints of the three slits on the sky are presented on the upper left, and the [O II] emission lines from the rectified 2D spectra obtained with the three slits are presented in the other three upper panels.  We use the {\sc rotcurve} program described in \cite{Weiner2006}, which takes into account the effects of seeing, to fit the emission lines. The rotation velocity and velocity dispersion as a function of radius are measured and presented in the middle and lower panels, respectively.  The solid points show the results of Gaussian fits to each row of the 2D spectra.  The black lines are the best-fit models to these points, and the red lines are the intrinsic models without the effect of seeing. We quantify the gas kinematics with two parameters, the rotation velocity measured at the flat part of the rotation curve and the intrinsic velocity dispersion, the latter representing the amount of disordered motions (\citealt{Kassin2007,Kassin2012,Covington2010, Stott2016}).  The galaxy rotates fastest, $61\pm 4\,\mathrm{km}\cdot\mathrm{s}^{-1}$, along the position angle at 10\degr, and slower at the other position angles, $41\pm 3\,\mathrm{km}\cdot\mathrm{s}^{-1}$ at 30\degr and $0\pm 4\,\mathrm{km}\cdot\mathrm{s}^{-1}$ at 58\degr, where a positive velocity indicates that the Arc region is redshifted with respect to the Central region.  
The intrinsic velocity dispersions are around $40\,\mathrm{km}\cdot\mathrm{s}^{-1}$, which are high compared to local disk galaxies but expected since there is likely a lot of disordered motions in this merger (e.g., the local galaxy NGC 4038 in the Antennae merging systems; \citealt{Ueda2012}).

In summary, an inspection of \emph{both} the gas kinematics and galaxy morphology leads to the conclusion that the galaxy is a major merger system or a disk severely distorted by a recent merger.  Interestingly, if the kinematics had to be interpreted without the aid of a resolved Hubble image, one might mistakenly perceive that the galaxy is an isolated and regularly rotating disk due to the limited spatial resolution of ground-based kinematics measurements (\citealt{Simons2019}).

\section{Co-adding Spectra for Each Region}

\label{sec:extractionspec}

Spectra are extracted from the Central and Arc regions demarcated on the \emph{HST} image of the galaxy in Figure \ref{fig:imageandregion}. The angular sizes of the regions are the same: 1.0\arcsec\ in the dispersion direction, and 0.8\arcsec\ in the spatial direction. Their size is comparable to that of  the seeing, which is 0.86\arcsec\ (FWHM). The two regions overlap by a small amount, no more than 20\% of the total area of each region.  As tested in Appendix \ref{appendix:beamsmearing}, the two regions are spatially distinct in our ground-based Keck observations.

In order to extract 1D spectra from these regions, we first spatially align each of the three 2D spectra (for the three position angles) to an \emph{HST} image.  The \emph{HST}/ACS F606W image is chosen because it has a similar wavelength range as the spectra.  For each spectrum, alignment is performed by matching the integrated flux profile of the galaxy with the flux profile inferred from the \emph{HST} image, where the flux is measured where the spectral slits overlap the regions, as shown in Figure \ref{fig:imageandregion}. The shape of the flux profiles measured from the spectra and the \emph{HST} image are remarkably similar.  Peaks of the profiles measured from the spectra and the \emph{HST} image are shifted in space to match each other in order to spatially align them.  Next, from each 2D spectrum which is taken with a certain slit position angle, we extract 1D spectra from the Central and Arc regions. As a demonstration of this process, the top panel of Figure \ref{fig:skysubtraction} shows where the extraction is made on an example 2D spectrum.

\begin{figure}
	\centering
	\includegraphics[width = 3.in]{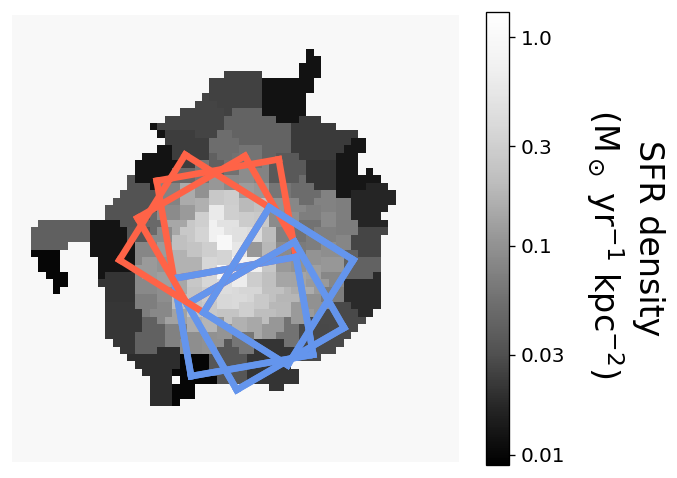}
	\caption{The SFR density map inferred from the SED fitting by de la Vega et al. in preparation. The Central and Arc regions are indicated in red and blue, respectively. The two regions have similar SFR densities: $0.23^{+0.26}_{-0.10}$ $M_\sun\,\mathrm{yr^{-1}kpc^{-2}}$ for the Central region and $0.15^{+0.04}_{-0.04}$ $M_\sun\,\mathrm{yr^{-1}kpc^{-2}}$ for the Arc region. 
		\label{fig:sfrdensity}}
\end{figure}

Finally, we combine the 1D spectra of the Central and Arc regions measured for each of the position angles.  They are shown as red and blue boxes in Figure \ref{fig:imageandregion}, respectively.  The resulting spectra are shown on the right side of the figure.  We fit the [\ion{O}{2}] line profiles in these spectra to measure the systemic redshift of each region. The redshifts are 1.3060 for the Central region and 1.3063 for the Arc region. More details about the fitting can be found in Appendix \ref{appendix:oiimeasurement}.

\section{Map of SFR Density from HST Images}

\label{sec:measuremasssfragn}

\begin{figure*}
	\centering
	\includegraphics[width = 7in]{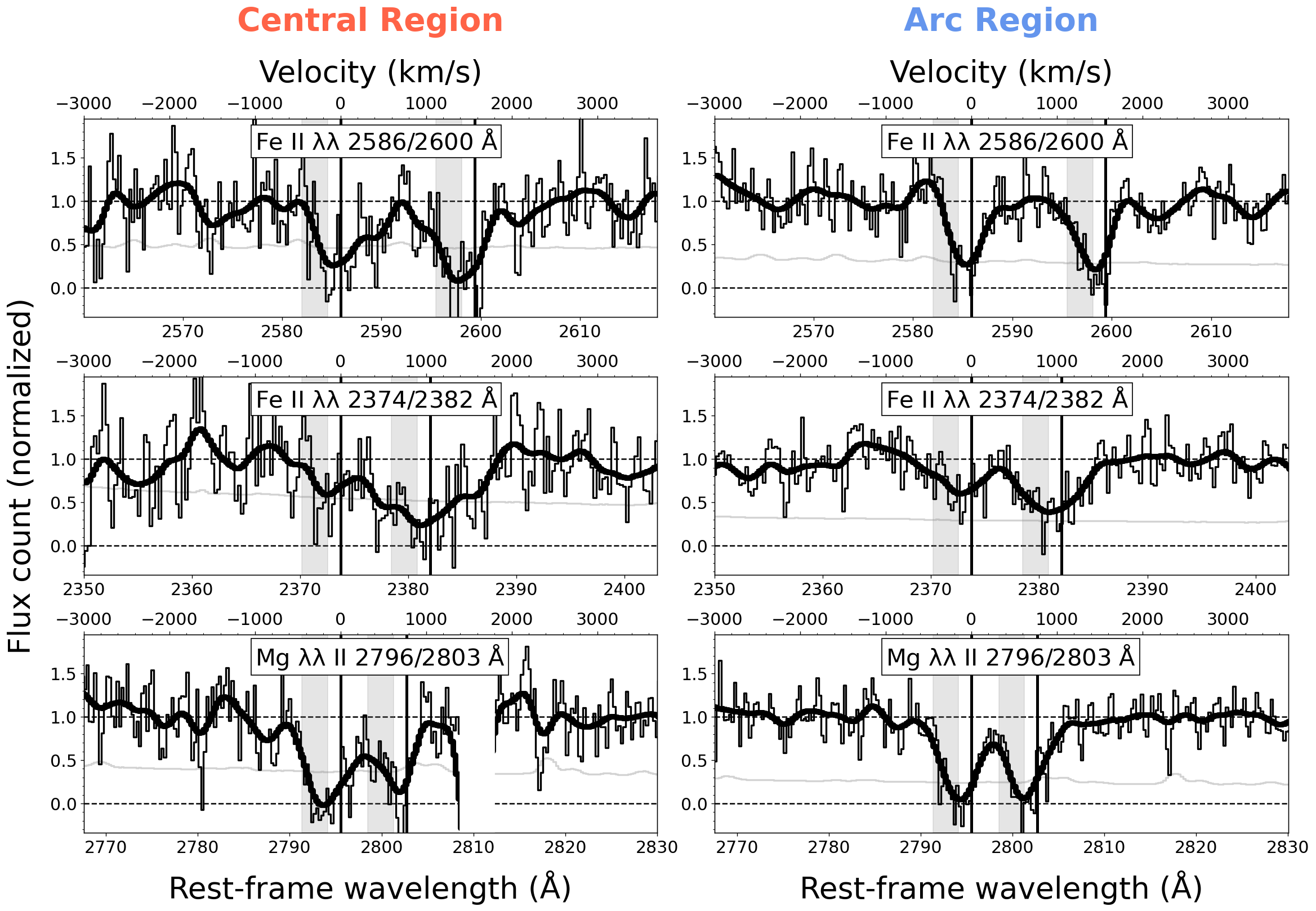}
	
	\caption{The line profiles of the \ion{Fe}{2} and \ion{Mg}{2} doublets are shown as thin black lines for the Central and Arc regions in the left and right columns, respectively.  The spectra have been shifted to the rest frame using the redshifts measured from the [\ion{O}{2}] emission lines.  These redshifts are taken as zero velocity and are indicated by the vertical lines.  Thick lines are smoothed versions of the spectra created by convolving them with a Gaussian with a standard deviation of 1.0 \AA\ in the rest frame.  The absorption line profiles of the two regions have similar shapes.  For both regions, the absorption line profiles extend to $\sim450$ $\mathrm{km}\cdot \mathrm{s}^{-1}$ blue-ward of zero velocity, which indicates cool gas outflows up to 450 $\mathrm{km}\cdot \mathrm{s}^{-1}$. The gray shading indicates the velocity range from which the line equivalent widths are measured in \S \ref{sec:ewmeasurement}, namely [--450, --150] $\mathrm{km}\cdot \mathrm{s}^{-1}$. The light gray lines represent the spectral flux uncertainties.
		\label{fig:absspec}}
	
\end{figure*}

We create a map of the SFR density by performing spatially resolved SED fitting of the {\it HST} images using the BEAGLE tool (\citealt{Chevallard2016}). To ensure that the mass-to-light ratio is reliably constrained, we bin pixels in the reddest \emph{HST} waveband, F160W, until they have a signal-to-noise of at least 10.  To do this, we adopt the Voroni binning algorithm of \cite{Cappelari2003}.  These bins are then applied to five other \emph{HST} wavebands: F435W, F606W, F850LP, F105W, and F125W.  For each spatial bin, we obtain the fluxes in all bands and perform SED fitting to them. The fitting assumes the initial mass function of \cite{Chabrier2003}, a delayed-exponential star-formation history, and the dust attenuation law by \cite{Charlot2000}  and \cite{Chevallard2013}. More details of our procedure are in de la Vega et al. in preparation. The inferred SFR density map is shown in Figure \ref{fig:sfrdensity}.

\begin{figure*}
	\centering
	\includegraphics[width = 7 in]{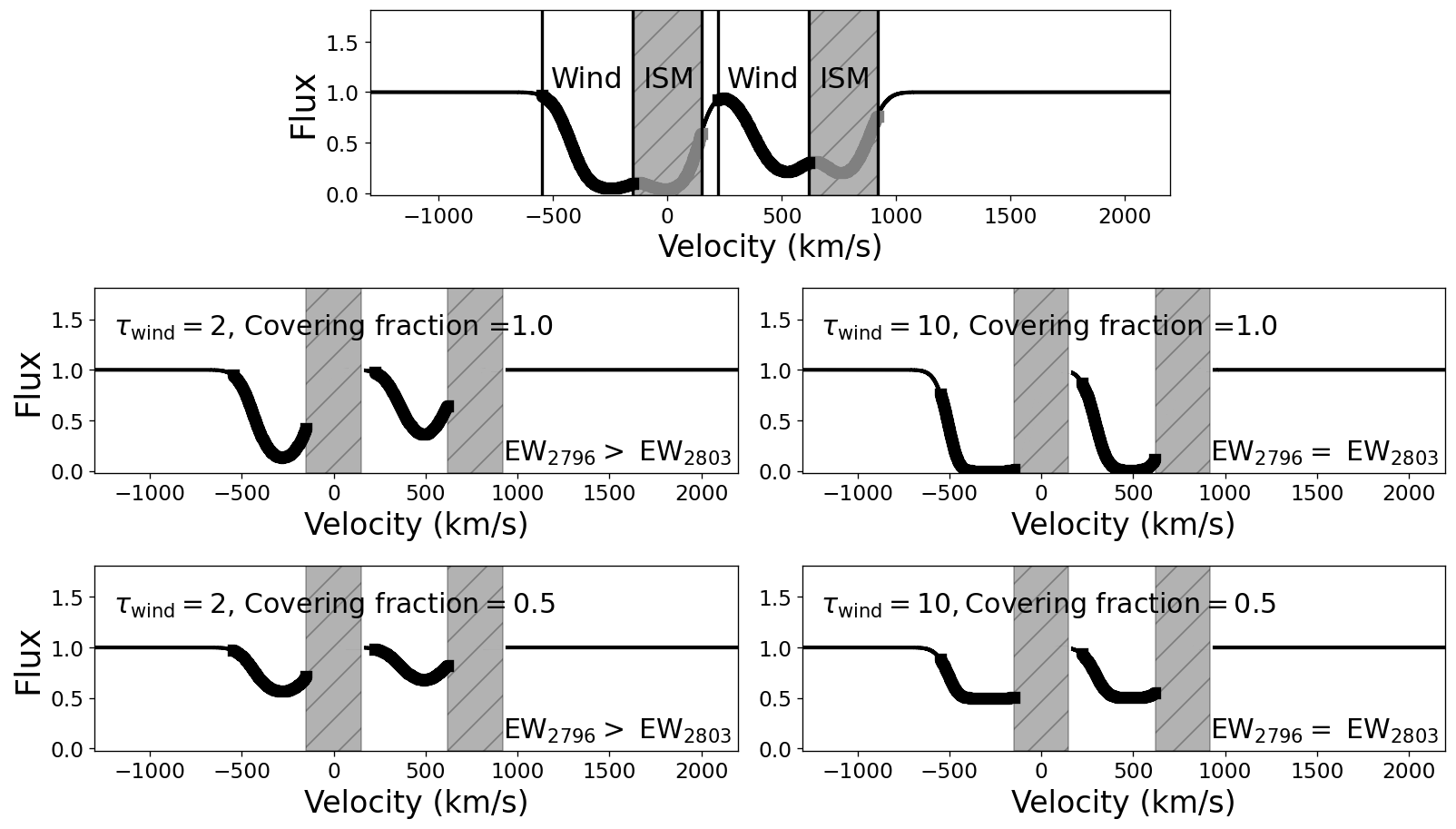}
	\caption{The column density of outflowing gas can be constrained by comparing the equivalent widths of absorption lines in a doublet. This is demonstrated using toy models of the \ion{Mg}{2} $\lambda\lambda$ 2796/2803 \AA\ doublet. \emph{Top panel:} In the rest frame, the ISM causes line absorption around zero velocity, and 
	the outflowing gas causes line absorption at negative velocities (moving toward the observer). The ISM component is only shown in this panel for clarity. \emph{Middle and bottom panels:} The line optical depth determines the differences between the equivalent widths of two lines in the doublet. This is demonstrated by models with different values of $\tau_\mathrm{wind}$, the optical depth at the line center of \ion{Mg}{2} 2796\,\AA, which are indicated in each panel. If  $\tau_\mathrm{wind}$ is low (left column), then the 2796 \AA\ line is deeper and has a larger equivalent width than the 2803 \AA\ line. As $\tau_\mathrm{wind}$ increases (right column), the 2803\,\AA\  line becomes stronger, and its profile and equivalent width become more similar to those of the 2796\,\AA\ line due to saturation. This is the case for any value of the gas covering fraction, which we show in the middle row and bottom row.  \label{fig:lineprofiledemo} }
\end{figure*}

Using the regions defined in Section \ref{sec:extractionspec} and demarcated in Figures \ref{fig:imageandregion} and \ref{fig:sfrdensity}, the SFR density of the Central region is measured to be $0.23^{+0.26}_{-0.10}\, M_\sun\,\mathrm{yr^{-1}kpc^{-2}}$, which is slightly higher than that of the Arc region, $0.15^{+0.04}_{-0.04}\, M_\sun\,\mathrm{yr^{-1}kpc^{-2}}$.   The SFRs are $18^{+19}_{-8}\, M_\sun \mathrm{yr}^{-1}$ and $13^{+4}_{-4}\, M_\sun \mathrm{yr}^{-1}$, respectively.  The star formation in the Central region is more obscured by dust ($A_\mathrm{V} = 1.0$) than that in the Arc region ($A_\mathrm{V} = 0.4$). The higher dust attenuation of the Central region is expected from its redder color as seen in the \emph{HST} image in Figure \ref{fig:sfms}.

\section{Properties of the cool outflowing gas from UV absorption lines}

\label{sec:measurewindline}

Properties of cool outflowing gas (around 10$^{4}$ K),  including velocities and column densities, are measured from the Central and Arc regions from the \ion{Fe}{2} and \ion{Mg}{2} absorption lines in this section.

\begin{figure*}
	\centering
	\includegraphics[width = 7. in]{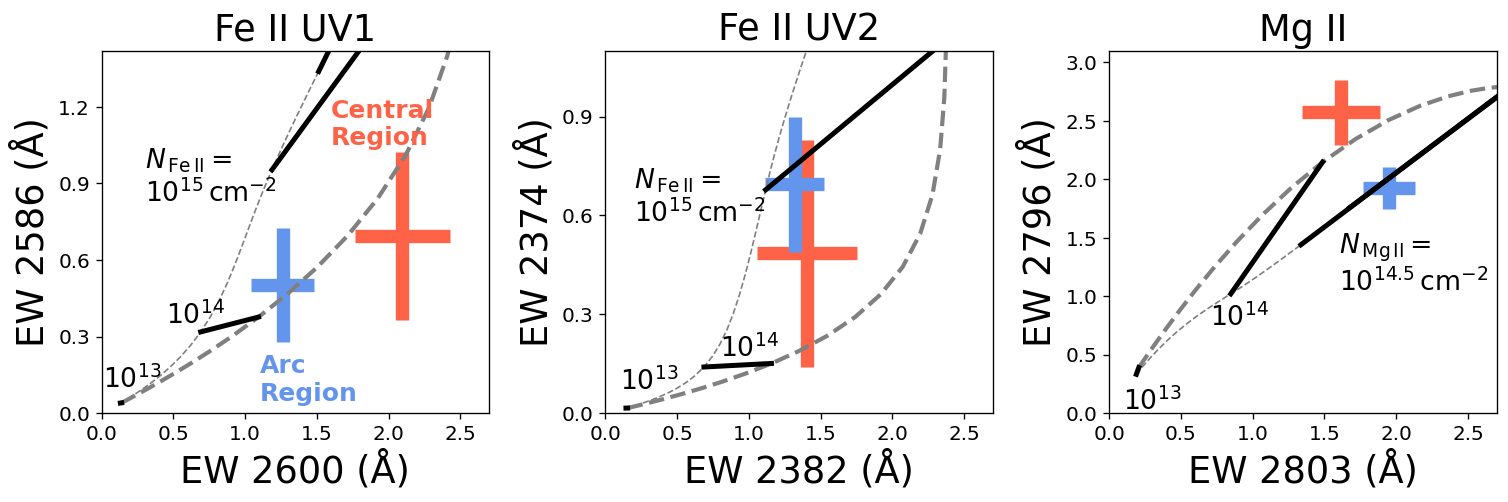}	
	\caption{The \ion{Fe}{2} and \ion{Mg}{2} column densities of outflowing gas are constrained by comparing the measured line equivalent widths with those calculated from simple models of absorption line profiles. From left to right, the equivalent widths of the \ion{Fe}{2} $\lambda\lambda$ 2586/2600 \AA, \ion{Fe}{2} $\lambda\lambda$ 2374/2382 \AA, and \ion{Mg}{2} $\lambda\lambda$ 2796/2803 \AA\ doublets are plotted. Red and blue crosses represent the measured equivalent widths for the Central and Arc regions, respectively. The models are shown as solid and dashed lines. Each solid line represents a series of models with the same column density, the value of which is given in the figure, but different velocity dispersions. Each dashed line represents a series of models with the same velocity dispersion but different column densities. The short and long dashed lines represent models with a velocity dispersion of 50 $\mathrm{km}\cdot \mathrm{s}^{-1}$ and 250 $\mathrm{km}\cdot \mathrm{s}^{-1}$, respectively. All the models assume that the gas covering fraction is unity.  Comparing the models with the observations, we infer \ion{Fe}{2} column densities of  $10^{14.0}$--$10^{15.0}\,\mathrm{cm}^{-2}$ for both regions.  Similarly, we infer a \ion{Mg}{2} column density of $10^{14.0}$--$10^{14.5}\,\mathrm{cm}^{-2}$ for the Central region and  $\gtrsim 10^{14.5}\,\mathrm{cm}^{-2}$ for the Arc region.
		\label{fig:ewdiagnostics}}
\end{figure*}

\subsection{Visual Inspection of the Absorption Line Profiles}

\label{sec:lineprofqualitativediscuss}

\begin{table}
	\renewcommand{\thetable}{\arabic{table}}
	\caption{Equivalent widths of the Fe and Mg lines for the Central and Arc regions.
		\label{tab:lineflambda}
	} 
	
	\centering
	\begin{tabular}{llll}
		\hline
		\hline
		Spectral line & $\lambda f$ (\AA)$^{a}$  & EW$_\mathrm{Central}$ (\AA)  & EW$_\mathrm{Arc}$ (\AA)  \\
		\hline
		\ion{Fe}{2} UV1 2586\,\AA &  178  & $0.7\pm 0.3$   & $0.5\pm 0.2$ \\
		\ion{Fe}{2} UV1 2600\,\AA & 624 & $2.1\pm 0.3$   & $1.3\pm 0.2$  \\
		\ion{Fe}{2} UV2 2374\,\AA &   73.6 & $0.5\pm 0.3$  & $0.7\pm 0.2$  \\
		\ion{Fe}{2} UV2 2382\,\AA &  762 &  $1.4\pm 0.3$  & $1.3\pm 0.2$  \\
		\ion{Mg}{2} 2796\,\AA &   1710     & $2.6\pm 0.3$  & $1.9\pm 0.2^{b}$ \\
		\ion{Mg}{2} 2803\,\AA &   869      &  $1.6\pm 0.3$  &  $1.9\pm 0.2^{b}$ \\
		Mg\,I 2852\,\AA & 5130 & $<1.1$ (3-$\sigma$) & $<0.7$ (3-$\sigma$)\\
		\hline
	\end{tabular}
	\begin{flushleft}
		\tablenotetext{a}{The product of the rest-frame wavelength $\lambda$ and the osciallator strength $f$ are from Table 2 of \cite{Zhu2015}.} 
		\tablenotetext{b}{The \ion{Mg}{2} doublet lines are fully saturated for the Arc region. Their equivalent widths are identical as a result.}
	\end{flushleft}
\end{table}

The line profiles of the \ion{Fe}{2} and \ion{Mg}{2} doublets are shown as thin black lines in Figure  \ref{fig:absspec}.  The spectra have been shifted to the rest frame using the redshifts measured from the [\ion{O}{2}] lines ($z=1.3060$ for the Central region and $z=1.3063$ for the Arc region), and normalized by the spectral continuum. We also show the smoothed spectra as thick black lines in this figure, which are generated by convolving the original spectra with a Gaussian with a standard deviation of 1.0 \AA\ in the rest frame. Two features of the line profiles can be identified via visual inspection:
 \begin{enumerate}
 	\item \emph{For both regions, all the absorption line profiles are  asymmetric, with wings extending to around $-450$ $\mathrm{km}\cdot \mathrm{s}^{-1}$}. This is indicative of outflowing gas moving away from the galaxy toward the observer (e.g., \citealt{Weiner2009}). 

	\item The absorption lines in each doublet most likely have different depths, which indicates that the doublet may not be fully saturated. The only exception is the \ion{Mg}{2} doublet from the Arc region for which the lines have nearly identical depths, indicating full saturation.  
	
 \end{enumerate}

We measure the equivalent widths of the \ion{Fe}{2} and \ion{Mg}{2} absorption lines in the following to compare the strengths of the lines quantitatively.

\subsection{Column Densities from Equivalent Widths}

\label{sec:ewmeasurement}

To constrain the column densities of \ion{Fe}{2} and \ion{Mg}{2} in the outflowing gas, we look into the equivalent widths of their absorption line doublets. The equivalent widths are measured within a velocity window of $[-450,\, -150]\,\mathrm{km}\cdot \mathrm{s}^{-1}$ relative to the systemic redshifts of each region. We select such a window for two reasons. First, it covers the maximum velocities where the blueshifted line absorptions are seen, as explained in \S \ref{sec:lineprofqualitativediscuss}. Second, it is offset from zero velocity by at least 2 times the ISM velocity dispersion values (around $65 \,\mathrm{km}\cdot \mathrm{s}^{-1}$, as measured from [\ion{O}{2}]; Appendix \ref{appendix:oiimeasurement}) such that the absorption caused by the gas of the interstellar medium (ISM) does not make a substantial contribution to the measured equivalent widths . Figure \ref{fig:lineprofiledemo} shows how the column density can be constrained by comparing the equivalent widths of two absorption lines in the same doublet.  Generally, a large difference in equivalent widths between the two lines indicates a low column density, and a nearly zero difference indicates a high column density.  A detailed explanation of the physics involved is provided in Appendix \ref{sec:ewversusn_physics} (see also \citealt{Draine2011}).

The measured equivalent widths are listed in Table \ref{tab:lineflambda} and plotted in Figure \ref{fig:ewdiagnostics}. In the figure, equivalent widths for the two \ion{Fe}{2} doublets and one \ion{Mg}{2} doublet are shown in three plots from left to right.  Red and blue crosses indicate measurements and their uncertainties for the Central and Arc regions, respectively.  The line with the shorter wavelength in each doublet is assigned to the vertical axis in each plot, and the scale is different on the vertical axes of the three plots.  For both regions, lines in the \ion{Fe}{2} doublets have different equivalent widths such that the line with the shorter wavelength has a smaller value.  Therefore, the \ion{Fe}{2} column densities in the two regions are not saturated. For \ion{Mg}{2}, the lines in the doublet have the same equivalent widths for the Arc region, indicating saturation.  The lines have different values for the Central region, with the line with the shorter wavelength having the higher equivalent width.  As per Figure~\ref{fig:lineprofiledemo}, this indicates that the \ion{Mg}{2} column density in the Central region is below a saturation value, whereas the \ion{Mg}{2} column density in the Arc region is saturated.

To provide quantitative constraints on the \ion{Fe}{2} and \ion{Mg}{2} column densities, we also compare in Figure \ref{fig:ewdiagnostics} the measured equivalent widths with those from simple analytic models of absorption line profiles.  Each model is generated from two input parameters: the integrated column density $N_{\mathrm{ion}}$ and the gas velocity dispersion $\sigma$.  Two steps are made to generate the models. First, the column density at a given line-of-sight velocity, $N_{v\mathrm{,\,ion}}(v)$, is calculated from the two input parameters by assuming that $N_{v\mathrm{,\,ion}}(v)$ is a Gaussian function that is centered at $-150$ km$\cdot$s$^{-1}$ and has a standard deviation of $\sigma$. Second, $N_{v\mathrm{,\,ion}}(v)$ is converted into the optical depth, $\tau_\mathrm{wind}(v)$, following \cite{Spitzer1978} and \cite{Arav2001}: 
\begin{equation}
	\tau_\mathrm{wind}(v) = 2.65\times 10^{-15}\, \lambda \, [ \mathrm{\AA} ]\, f N_{v\mathrm{,\,ion}}(v) \ \mathrm{[cm^{-2}\,km^{-1}\,s]},
	\label{eqn:tauversusfN}
\end{equation}
where  $\lambda$ is the rest-frame wavelength of the absorption line and $f$ is the quantum oscillator strength. The adopted values of $\lambda f$ are in Table \ref{tab:lineflambda}.  Finally, absorption line profiles are inferred from $\tau_\mathrm{wind}(v)$ by assuming that the outflowing gas fully covers the galaxy. The line flux density normalized by the spectral continuum, $F(v)$, is calculated as $F(v) = e^{-\tau_\mathrm{wind}(v)}$.

After the model line profiles are constructed, equivalent widths are measured from them in the same velocity window as for the observations. The equivalent widths are plotted as solid and dashed lines in Figure \ref{fig:ewdiagnostics}. Each solid line represents a series of models with the same column density, whose numerical value is indicated in the figure, but different velocity dispersions. There are two dashed lines which represent a series of models with the same velocity dispersion but different column densities. The short dashed lines show models with a velocity dispersion of 50$\,\mathrm{km}\cdot \mathrm{s}^{-1}$, and the long dashed lines show models with 250$\,\mathrm{km}\cdot \mathrm{s}^{-1}$.

The models match observations fairly well in Figure \ref{fig:ewdiagnostics}. By comparing the observations with the models, we can infer the \ion{Fe}{2} column densities of the two regions to be between $10^{14.0}$ and $10^{15.0}\,\mathrm{cm}^{-2}$.  The \ion{Mg}{2} column density for the Arc region is inferred to be $\gtrsim 10^{14.5}\,\mathrm{cm}^{-2}$, and that for the Central region is between $10^{14.0}$ and $10^{14.5}\,\mathrm{cm}^{-2}$.  Statistical uncertainties for these values are difficult to infer from Figure \ref{fig:ewdiagnostics}. However, they can be inferred from a more sophisticated method that we use in \S\ref{sec:functionfittinglines}.

\subsection{Covering Fractions \& Column Densities from Line Profile Fitting}

\label{sec:functionfittinglines}

\begin{figure*}
	\centering
	\includegraphics[width = 7. in]{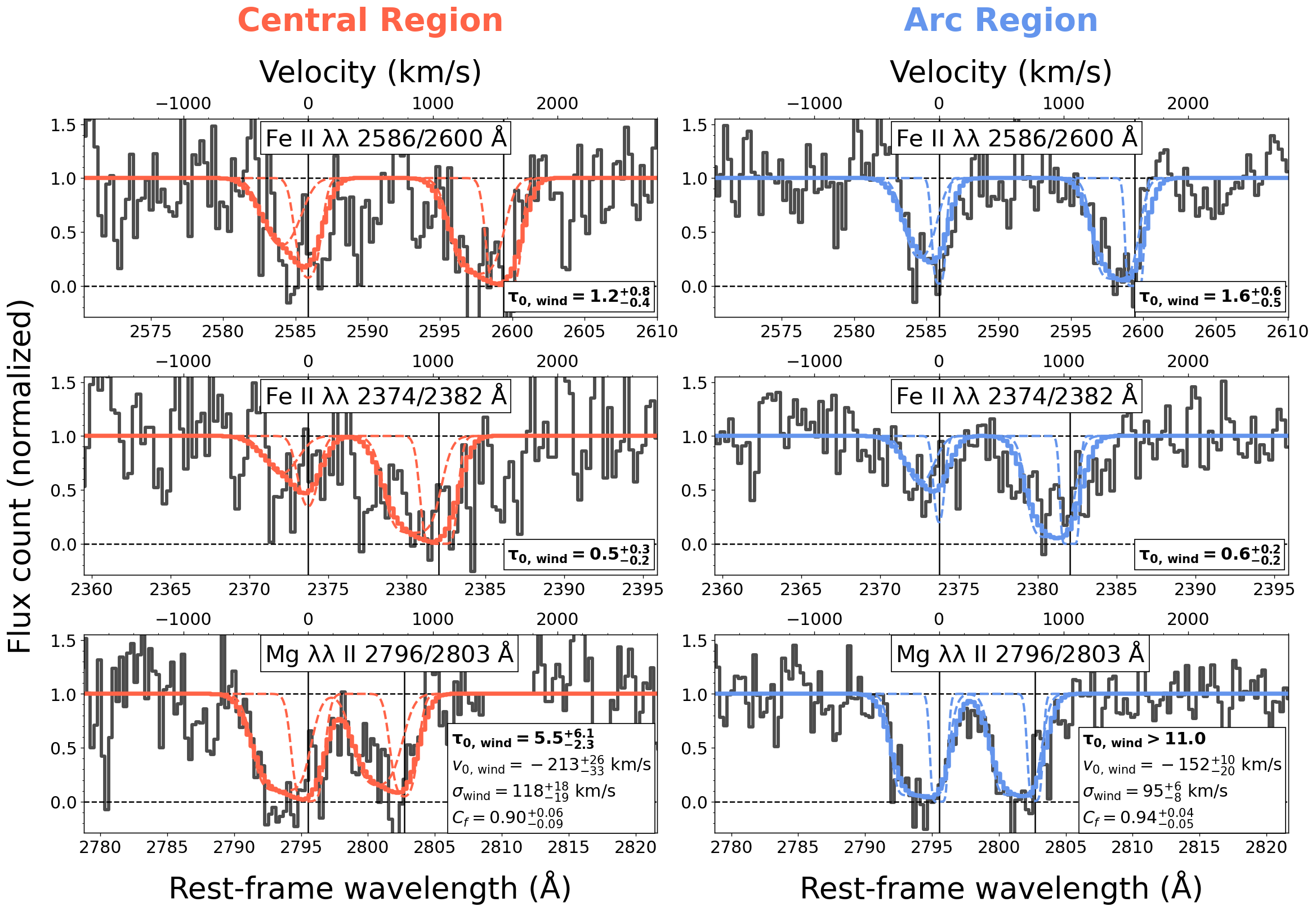}
	
	\caption{The optical depths and covering fractions of winds are measured by fitting absorption line profiles (black lines) with simple analytic models.  The fits for the Central and Arc regions are shown in the left and right columns, and are colored red and blue, respectively.  The \ion{Fe}{2} doublets are shown in the top two rows, and the \ion{Mg}{2} doublet is in the bottom row.  Vertical lines indicate zero velocity in the rest-frame. The fitting incorporates two components, one for the ISM and one for the wind, both of which are shown as thin dashed lines. Most probable values of the free parameters in the fits are given in the figure and Table  \ref{tab:lineprofilefitresult}.  Winds in the two regions have similar values of $\tau_{0,\,\mathrm{wind}}$ for the \ion{Fe}{2} lines.  The $\tau_{0,\,\mathrm{wind}}$ values for \ion{Mg}{2} are likely higher in the Arc region than the Central region. The gas covering fractions are greater than 0.9 for both regions.  \label{fig:tau_and_cf}}

\end{figure*}

To provide more quantitative constraints on the gas covering fractions and ion column densities, we  perform a simultaneous fit to the observed \ion{Fe}{2} and \ion{Mg}{2} absorption line profiles using an analytic model.  These line profiles are shown in Figure \ref{fig:tau_and_cf} as black lines, where the left column shows the profiles for the Central region and the right for the Arc region.

\begin{table*}
	\renewcommand{\thetable}{\arabic{table}}
	\caption{Properties of the outflowing gas and ISM measured from fitting the \ion{Fe}{2} and \ion{Mg}{2} line profiles for the Central and Arc regions} \label{tab:lineprofilefitresult}
	\centering
	\begin{tabular}{lll}
		\hline
		\hline
		Quantity & Central region & Arc region  \\
		\hline 
		\bf{Outflowing gas:} & & \\
		Central optical depth $\tau_\mathrm{0,\,\mathrm{wind}}$ of \ion{Fe}{2}\,2586\,\AA\ & $1.2^{+0.8}_{-0.4}$ & $1.6^{+0.6}_{-0.5}$ \\
		Central optical depth $\tau_\mathrm{0,\,\mathrm{wind}}$ of \ion{Fe}{2}\,2374\,\AA\ & $0.5^{+0.3}_{-0.2}$ & $0.6^{+0.2}_{-0.2}$ \\
		Central optical depth $\tau_\mathrm{0,\,\mathrm{wind}}$ of \ion{Mg}{2}\,2796\,\AA\ & $5.5^{+6.1}_{-2.3}$ & $>11.0$ \\
		Central velocity $v_\mathrm{0,\,\mathrm{wind}}$ & $-213^{+26}_{-33}\ \mathrm{km}\cdot \mathrm{s}^{-1}$ & $-152^{+10}_{-20}\ \mathrm{km}\cdot \mathrm{s}^{-1}$  \\
		Velocity dispersion $\sigma_\mathrm{wind}$ & $118^{+18}_{-19}\ \mathrm{km}\cdot \mathrm{s}^{-1}$ & $95^{+6}_{-8}\ \mathrm{km}\cdot \mathrm{s}^{-1}$  \\
		Gas covering fraction $C_f$ &  0.90$^{+0.06}_{-0.09}$ & 0.94$^{+0.04}_{-0.05}$ \\
		$\log$ \ion{Fe}{2} column density$^a$  $\log\,N_\mathrm{Fe\,II}$ &      $14.7^{+0.2}_{-0.2}\ \mathrm{cm}^{-2}$                     &                   $14.6^{+0.1}_{-0.2}\ \mathrm{cm}^{-2}$                         \\
		$\log$ \ion{Mg}{2} column density$^a$ $\log\,N_\mathrm{Mg\,II}$ &          $14.4^{+0.3}_{-0.2}\ \mathrm{cm}^{-2}$                 &                  $>14.5\ \mathrm{cm}^{-2}$                        \\
		\bf{ISM:} & & \\
		Central optical depth $\tau_\mathrm{0,\,\mathrm{ISM}}$ of \ion{Fe}{2}\,2586\,\AA\ & $2.6^{+2.9}_{-1.3}$ & $3.9^{+6.0}_{-2.3}$ \\
		Central optical depth $\tau_\mathrm{0,\,\mathrm{ISM}}$ of \ion{Fe}{2}\,2374\,\AA\ & $1.1^{+1.2}_{-0.5}$ & $1.6^{+2.5}_{-1.0}$ \\
		Central optical depth $\tau_\mathrm{0,\,\mathrm{ISM}}$ of \ion{Mg}{2}\,2796\,\AA\ & $6.0^{+7.2}_{-3.2}$ & $9.9^{+6.8}_{-6.6}$ \\
		Velocity dispersion $\sigma_\mathrm{ISM}$ & $67^{+18}_{-14}\ \mathrm{km}\cdot \mathrm{s}^{-1}$ &  $33^{+12}_{-11}\ \mathrm{km}\cdot \mathrm{s}^{-1}$ \\
		\hline
	\end{tabular}
	
	\begin{flushleft}
	
		\tablenotetext{a}{These are calculated over the velocity range $-450$ $\mathrm{km}\cdot \mathrm{s}^{-1}$ to $-150$ $\mathrm{km}\cdot \mathrm{s}^{-1}$.}	
	\end{flushleft}
	
\end{table*}

\subsubsection{Fitting Methodology}

For fits to the line profiles, we adopt a functional form that reflects absorption from winds and the ISM. For each component, the line optical depth $\tau$ is assumed to be a Gaussian function of the line-of-sight velocity $v$. Gas from the ISM is assumed to fully cover the galaxy along the lines-of-sight, whereas the wind component has a covering fraction $C_f$ that ranges between 0.0 and 1.0. The resulting line profiles as a function of velocity, $F(v)$, are determined by the optical depths due to the wind and the ISM,  $\tau_\mathrm{\,wind}(v)$ and  $\tau_\mathrm{ISM}(v)$, and $C_f$.  A continuum-normalized line profile has the following shape:
\begin{equation}
F(v) =  [1-C_f+C_f\cdot e^{-\tau_\mathrm{wind}(v)}] \cdot e^{-\tau_\mathrm{\,ISM}(v)}. \label{eqn:fluxprofile}
\end{equation}
The bracketed term on the right side represents absorption by the wind.  The second term represents absorption by the ISM. The optical depths are:
\begin{equation}
\tau_\mathrm{wind}(v) = \tau_{0,\, \mathrm{wind}}\cdot e^{-(v-v_{0,\,\mathrm{wind}})^2/(2\sigma_\mathrm{wind}^2)},
\label{eqn:lineformwind}
\end{equation}
and
\begin{equation}
\tau_\mathrm{\,ISM}(v) = \tau_\mathrm{0,\ ISM}\cdot e^{-v^2/(2\sigma_\mathrm{\,ISM}^2)},
\label{eqn:lineformism}
\end{equation}
where  $ \tau_\mathrm{0,\, wind}$ and $\tau_\mathrm{0,\, ISM}$ are optical depths at the central wavelengths of the wind and ISM components, respectively, $v_\mathrm{0,\, wind}$ is the central wind velocity, and $\sigma_\mathrm{wind}$ and $\sigma_\mathrm{\,ISM}$ are the velocity dispersions of the wind and the ISM, respectively.  Our modeling of the \ion{Fe}{2} and \ion{Mg}{2} line profiles does not include any component for line emission. This is because no \ion{Fe}{2}, \ion{Fe}{2}*, or \ion{Mg}{2} emission lines are detected in the spectra for the two regions, as shown in Figure \ref{fig:imageandregion}. We discuss possible reasons for the missing emission lines in \S \ref{sec:missingemissionline}.

Next, Equation \ref{eqn:fluxprofile} for the line profile shape is convolved with the line spread function of our observations ($70\,\mathrm{km}\cdot \mathrm{s}^{-1}$), and is then fit to the observed absorption line profiles. The fitting is performed with the Markov chain Monte Carlo software package, {\sc{emcee}} (\citealt{Foreman-Mackey2013,Foreman-Mackey2019}). Photometric uncertainties in the line profiles are taken into account in the fitting.

In the fitting, the following quantities are assumed to be the same for lines \emph{from the same spatial region}: $C_f$, $v_{0,\,\mathrm{wind}}$,  $\sigma_\mathrm{wind}$, and $\sigma_\mathrm{\,ISM}$. In addition, for lines \emph{associated with the same chemical element and from the same region}, their optical depths ($\tau_\mathrm{0,\, ISM}$ and $\tau_\mathrm{0,\, wind}$) are set to be proportional to each other. The ratios between their optical depths are equal to the ratios between their $\lambda f$ values, as in Equation \ref{eqn:tauversusfN}. The $\lambda f$ values of the \ion{Fe}{2} and \ion{Mg}{2} absorption lines are listed in Table \ref{tab:lineflambda}.

In the fits, uniform priors are assumed for the gas covering fraction: $0\leq C_f\leq 1$. Log-linear priors are assumed for the line optical depths: $0.01<\tau_\mathrm{0,\,wind}<20$ and $0.01<\tau_\mathrm{0,\,ISM}<20$, such that the probability of a given value of $\tau$ is a constant function of $\log\,\tau$. The central velocities of the winds are assigned uniform priors: $-350\ \mathrm{km}\cdot \mathrm{s}^{-1} <v_0<-50\ \mathrm{km}\cdot \mathrm{s}^{-1}$. Velocity dispersions of the ISM and wind components also have uniform priors: from 0 to 100 $\mathrm{km}\cdot \mathrm{s}^{-1}$ and from 0 to 250 $\mathrm{km}\cdot \mathrm{s}^{-1}$, respectively.

For the free parameters in Equations 2--4 ($C_f$, $\tau_{0,\,\mathrm{wind}}$, $\tau_{0,\,\mathrm{ISM}}$, $v_\mathrm{0,\, wind}$, $\sigma_\mathrm{0,\, wind}$, and $\sigma_\mathrm{0,\, ISM}$), the 50th percentiles of their posterior distributions are reported as the ``most probable" values.  These are given in Table \ref{tab:lineprofilefitresult}.  The corresponding ``best fit" models are shown in Figure 8 as red lines for the Central region and blue lines for the Arc region.  The 16th and 84th percentiles quantify the uncertainties of the free parameters.  The only exception is the central optical depth, $\tau_\mathrm{0,\,wind}$, of the \ion{Mg}{2} 2586\,\AA\ line for the Arc region. Its posterior distribution is skewed toward the upper bound of the prior, which is caused by the saturation of the \ion{Mg}{2} doublet in the Arc region (Figure \ref{fig:tau_and_cf}). As a result, only a lower limit can be inferred for its central optical depth, which is calculated as the 16th percentile of its posterior.

\begin{figure*}
	\centering
	\includegraphics[width = 7.2 in]{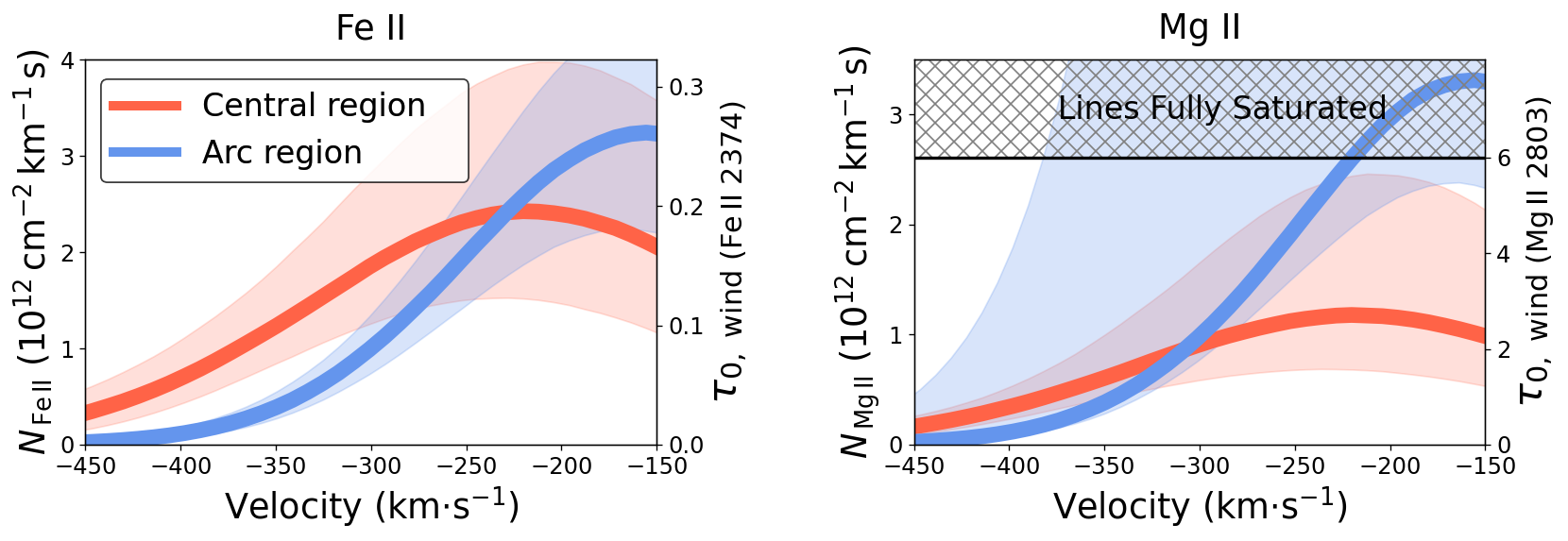}
	\caption{The \ion{Fe}{2} and \ion{Mg}{2} column densities of the outflowing gas as a function of velocity are shown in the left and right panels, respectively. The hatched box on the right indicates the regime where the \ion{Mg}{2} lines are fully saturated.  The column density values for the Central and Arc regions are shown in red and blue, respectively,in each panel. These values are the results of the line profile fitting in \S \ref{sec:functionfittinglines}.  Thick solid lines demarcate the 50th percentiles, and the lighter shading around them the 16th and 84th percentiles. The two regions have similar \ion{Fe}{2} column density profiles. The \ion{Mg}{2} column density profiles of the two regions are difficult to compare, since the \ion{Mg}{2} lines of the Arc region are saturated and no upper bound can be obtained for the corresponding column density at $v>-400\,\mathrm{km}\,\mathrm{s}^{-1}$.
	\label{fig:nionprof}}		
\end{figure*}

\subsubsection{Results of the Fitting}

\label{sec:mcmcresult}

The most probable line profile models from the fitting are shown in Figure \ref{fig:tau_and_cf} as thick red and blue lines, which show good matches to data (black lines). These models are generated using the most probable values of the parameters described above, including $C_f$, $\tau_{0,\,\mathrm{wind}}$, $\tau_{0,\,\mathrm{ISM}}$, $v_\mathrm{0,\, wind}$, $\sigma_\mathrm{0,\, wind}$, and $\sigma_\mathrm{0,\, ISM}$. These values and the corresponding uncertainties are listed in the lower right of each panel in Figure \ref{fig:tau_and_cf} and in Table \ref{tab:lineprofilefitresult}. 

For the ISM component, we note that the velocity dispersions inferred from the absorption lines, $67^{+18}_{-14}\ \mathrm{km}\cdot\mathrm{s}^{-1}$ for the Central region and $33^{+12}_{-11}\ \mathrm{km}\cdot\mathrm{s}^{-1}$ for the Arc region, are consistent with those inferred from the \ion{O}{2} emission lines, which are 30-50 $\mathrm{km}\cdot\mathrm{s}^{-1}$ (\S \ref{sec:morphology}).

For the wind component, the inferred gas covering fractions are high for the Central and Arc regions, 0.90 and 0.94, respectively. 

For the wind component, we also calculate the \ion{Fe}{2} and \ion{Mg}{2} column densities as a function of the line-of-sight velocity from the line profile fitting results using the following equation, which is a rearrangement of Equation \ref{eqn:tauversusfN}:
\begin{equation}
N_{v\mathrm{,\,ion}}(v)\ \mathrm{[cm^{-2}\,km^{-1}\,s]} = \frac{3.77\times 10^{14}}{\lambda \mathrm{[\AA]}f} \tau_\mathrm{wind}(v). 
\label{eqn:columndensityprof}
\end{equation}
The $\lambda f$ value of each line is listed in Table \ref{tab:lineflambda}, and $\tau_\mathrm{wind}(v)$ is calculated from Equation \ref{eqn:lineformwind} using the most probable values of $\tau_\mathrm{0,\,wind}$, $v_\mathrm{0,\,wind}$, and $\sigma_\mathrm{0,\,wind}$ which are listed in Table \ref{tab:lineprofilefitresult}. The calculated column densities as a function of velocity are shown in Figure \ref{fig:nionprof}, with the Central region in red and the Arc region in blue.  The two  regions have similar \ion{Fe}{2} column density profiles. The \ion{Mg}{2} column density profiles of the two regions are difficult to compare, since the \ion{Mg}{2} lines of the Arc region are saturated and no upper bound can be obtained for the corresponding column density.

Finally, we calculate the total column densities of \ion{Fe}{2} and \ion{Mg}{2} of the outflowing gas, $N_\mathrm{ion}$, by integrating the term $N_{v,\mathrm{\,ion}} (v)$ from Equation \ref{eqn:columndensityprof} over a velocity range from $-450$ to $-150$ km$\cdot$s$^{-1}$. Values of the \ion{Fe}{2} column densities are $14.7^{+0.2}_{-0.2}\ \mathrm{cm}^{-2}$ and                   $14.6^{+0.1}_{-0.2}\ \mathrm{cm}^{-2}$ for the Central and Arc regions, respectively, and values of the \ion{Mg}{2} column densities are $14.4^{+0.3}_{-0.2}\ \mathrm{cm}^{-2}$ and $>14.5\ \mathrm{cm,}^{-2}$ respectively. These values are also listed in Table \ref{tab:lineprofilefitresult}. Their typical uncertainties are around 0.3 dex. The \ion{Fe}{2} column densities of the two regions are similar within 0.5 dex, whereas the \ion{Mg}{2} column densities are difficult to compare since only a lower limit is obtained for the Arc region. These values are consistent with those measured from equivalent widths in \S \ref{sec:ewmeasurement}.

\begin{figure*}
	\centering
	\includegraphics[width = 6.5 in]{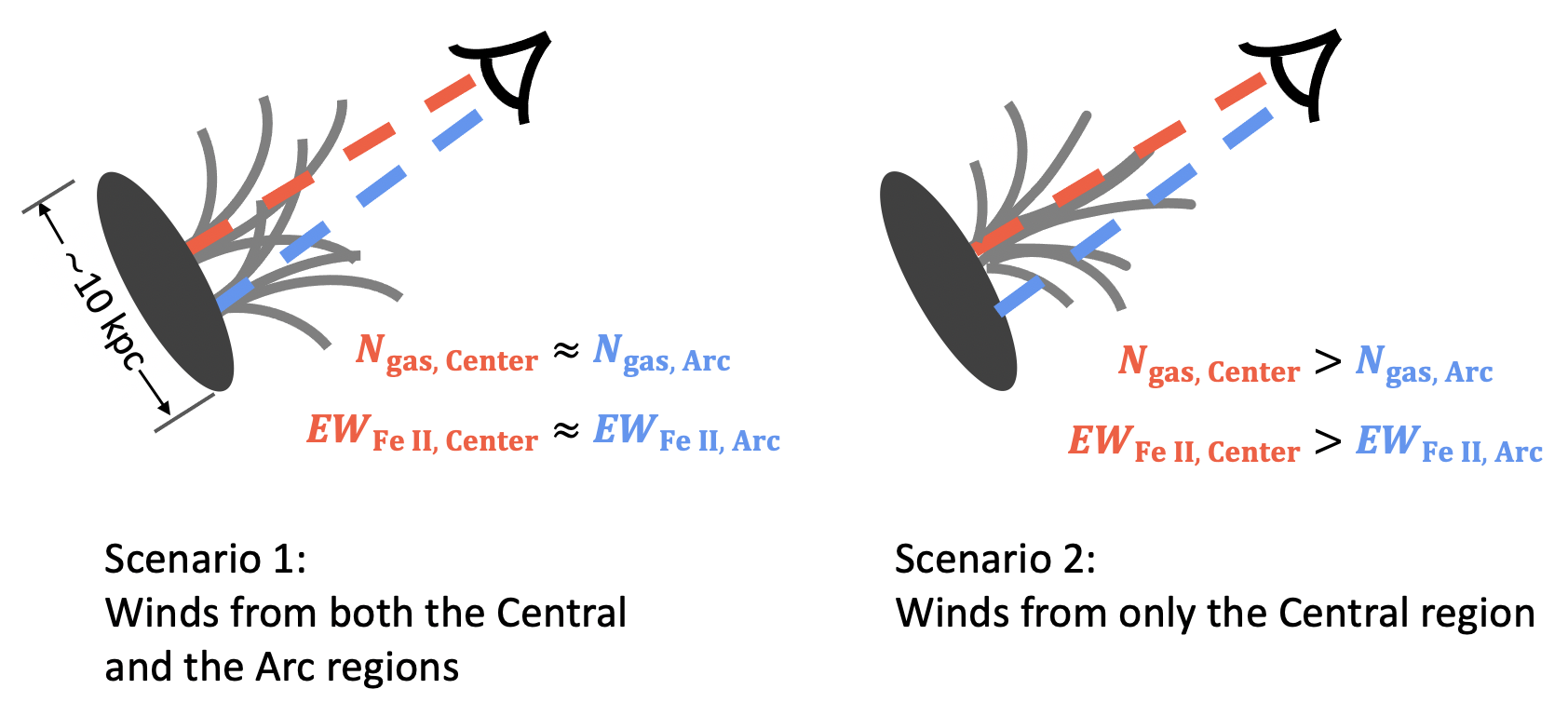}
	\caption{Two scenarios for wind launch locations are illustrated.  \emph{Left:} For Scenario 1, winds are launched from both the Central and Arc regions. They are assumed to have similar velocity and mass outflow rates. The gas column densities along the sightlines toward the two regions are also similar.  Therefore, the absorption line equivalent widths will be similar. \emph{Right:} For Scenario 2, winds are launched exclusively from the center of the galaxy but are observed from the same two sightlines as in Scenario 1. The line equivalent width is larger along the sightline toward the Central region,  because it passes through greater amount of outflowing gas. The column density profiles will be different for the two sightlines. Scenario 1 is favored in this work according to the observations, because the \ion{Fe}{2} column density profiles for the two regions are similar as shown in Figure \ref{fig:nionprof}. \label{fig:cartoon}}
\end{figure*}

\section{The Missing \ion{F\lowercase{e}}{2}* and \ion{M\lowercase{g}}{2} Emission Lines}

\label{sec:missingemissionline}

In a model where the outflowing gas has an isotropic distribution around the galaxy and is free of dust, the absorption and emission line associated with the same ion and upper energy level are expected to have comparable strengths \citep{Prochaska2011}.  Given that we observe strong \ion{Fe}{2} and \ion{Mg}{2} absorption lines, we should expect to find comparably strong \ion{Fe}{2}, \ion{Fe}{2}* and \ion{Mg}{2} \emph{emission} lines, including \ion{Fe}{2} 2586\,\AA, \ion{Fe}{2} 2600\,\AA, \ion{Fe}{2}* 2365\,\AA,  \ion{Fe}{2}* 2396\,\AA, \ion{Fe}{2}* 2612\,\AA, \ion{Fe}{2}* 2626\,\AA, \ion{Fe}{2}* 2631\,\AA, \ion{Mg}{2} 2796\,\AA,  and \ion{Mg}{2} 2803\,\AA,  the wavelengths of which are indicated by the blue tick marks in Figure \ref{fig:imageandregion}.
None of these emission lines are detected in the observations. 

It remains an open question why the emission lines are absent. Below we discuss three possible reasons. These explanations remain to be examined quantitatively in the future through radiative transfer modeling and deep observations with wide-field integral field spectrographs.

One possible reason is that the spatial extent of winds is significantly larger than the size of the galaxy itself (e.g., \citealt{Wang2020}).  In this case, the emission originates from a region outside of the two spatial regions in \S \ref{sec:measurewindline}.  We searched for these lines in a spectrum extracted from the full length of the slit, which has a length of 10.8,\arcsec but did not detect them.   However, it could be that the line emission has low enough surface brightness that it is lost beneath the sky background \citep{Prochaska2011}. Two other possible culprits are the dust in the outflowing gas and an anisotropic gas distribution.  Dust attenuates the emission line photons, making them to faint to detect.  An anisotropic gas distribution can cause photons to be re-directed from their paths along the line of sight, also making the lines too faint to detect.  

These two explanations are also proposed by several other observational studies at $z\sim 1$ (e.g., \citealt{Erb2012,Kornei2012,Kornei2013,Finley2017,Finley2017a,Feltre2018,RickardsVaught2019}).  Furthermore, observational studies also find that the \ion{Mg}{2} emission lines are present preferentially in galaxies with stellar masses below $10^{10}\,M_\sun$ and relatively low dust attenuation ($A_\mathrm{UV}<2$) (\citealt{Erb2012,Kornei2012,Kornei2013, Zhu2015,Finley2017a,Feltre2018,Henry2018}). The Baltimore Oriole's Nest galaxy has a higher stellar mass than these \ion{Mg}{2} emitters, although its $A_\mathrm{UV}$, which is estimated to be around 1.2 (\citealt{Pacifici2012,Pacifici2016}),  is within the range of the emitters. The \ion{Fe}{2}* emission lines are found preferentially in galaxies with stellar masses above $10^{10}\,M_\sun$ and $A_\mathrm{UV}<2$ (\citealt{Erb2012,Kornei2013,Finley2017a}). The Baltimore Oriole's Nest galaxy falls within the same mass and dust attenuation ranges but has no significant  \ion{Fe}{2}* lines, which remains to be understood by future studies through detailed radiative transfer modeling.

\section{Where are the winds launched from?}

\label{sec:discusswindorigin}

We consider two scenarios for where the winds are launched: from the Central and Arc regions (Scenario 1), and only from the Central region (Scenario 2) which is the case for the local galaxy M82 (\citealt{Lehnert1999,Heckman2017}).  They are illustrated in the left and right panels of Figure~\ref{fig:cartoon}, respectively.  In the figure, two sightlines are indicated, one toward the Central region and the other toward the Arc region. The winds are indicated with gray curves. 

Scenario 1 is favored because it naturally explains why the \ion{Fe}{2} column densities of winds observed along the sightlines toward the Central and Arc regions are similar (\S\ref{sec:measurewindline}). This is because comparable amounts of gas are launched from both regions of the galaxy, as indicated in Figure \ref{fig:cartoon}.   

In contrast, Scenario 2 struggles to explain the  similar \ion{Fe}{2} column densities observed along the sightlines toward the two regions (Figure \ref{fig:nionprof} and Table~\ref{tab:lineflambda}).  This scenario predicts that the column density should decline with radius from the center of the galaxy, or equivalently, the column density should be higher in the Central region than in the Arc region, which is not seen in the observations. To demonstrate this point, below we use a simple model to quantify the column densities of the two regions predicted by Scenario 2.

In the model, we assume that the wind is launched from the galaxy center with a density profile $n(r) \propto r^{-1}$, where $n$ is the volume density of gas and $r$ is the radial distance in kiloparsecs \citep{Burchett2020,Wang2020}. We further adopt a radial velocity profile of the wind ($v(r)$) which increases with $r$ \citep{Wang2020}, $v(r)\propto r$.  We define a quantity $r_\mathrm{max}$, the radial distance where the observed maximum wind velocity, 450 km$\cdot$s$^{-1}$ (\S \ref{sec:lineprofqualitativediscuss}), is reached. The value of $r_\mathrm{max}$ is assumed to be 15 kpc, consistent with the maximum radial extent of the cool galactic winds found by recent observations \citep{Burchett2020,Zabl2021}. With $n(r)$ and $v(r)$, we are able to calculate the column densities of the wind along the sightlines toward the Central region, which has an impact parameter of 0 kpc, and the Arc region, which has an impact parameter of 7 kpc, by integrating the volume density of gas along the two sightlines \citep{Sobolev1960}. At a line-of-sight velocity of $-150$ km$\cdot$s$^{-1}$, which is approximately the central velocity of the wind components in Figure \ref{fig:tau_and_cf}, the column density calculated for the Central region is 1.7 times that for the Arc region. This is in tension with the \ion{Fe}{2} column densities (Figure \ref{fig:nionprof}), where the column density for the Central region is $0.8^{+0.9}_{-0.4}$ (80\% confidence interval) times that of the Arc region at the same line-of-sight velocity. 

The functional forms of the gas density and velocity profiles described above are adopted because they are consistent with recent observational constraints on the structure of the cool galactic winds (e.g., \citealt{Wang2020,Burchett2020,Zabl2021}). Adopting alternative forms only makes Scenario 2 (wind exclusively from the Central region) even less plausible.  To demonstrate this, we consider an alternate form of the velocity profile seen in some recent simulations,  $v(r)\propto \sqrt{r}$ \citep{Schneider2020,Hopkins2021}, and a steep density profile seen in the simulation by \cite{Schneider2020}, $n(r) \propto r^{-2}$. If either or both of the profiles are adopted, the \ion{Fe}{2} column density of the Central region calculated for Scenario 2 is more than 3 times that of the Arc region,  which is in tension with observations at a confidence level above 95\%.

\begin{table*}
 	\caption{Assumed values of physical parameters for the calculation of mass outflow rates \label{tab:assumptions}}
 	\centering
 	\begin{tabular}{llll}
 	\hline \hline 
 	Parameter &  Description & Adopted value & Equation \\ \hline
 	$N_\mathrm{Fe\,II}$/$N_\mathrm{Fe, \,gas}$       &\ion{Fe}{2} ionization fraction      &   1.0             &  \ref{eqn:ionizationcorrection} \\
 	$N_\mathrm{Mg\,II}/N_\mathrm{Mg,\,gas}$          &\ion{Mg}{2} ionization fraction      &   1.0             &  \ref{eqn:ionizationcorrection} \\
 	$\log$ (Fe/H)$_\mathrm{total}$                   &Fe total abundance (gas+dust)        &    --4.5 (solar)  &  \ref{eqn:gasabundance}          \\
 	$\log$ (Mg/H)$_\mathrm{total}$                   &Mg total abundance (gas+dust)        &     --4.4 (solar) &  \ref{eqn:gasabundance}          \\
 	$[X_\mathrm{Fe}/\mathrm{H}]$                              &Fe dust depletion factor (log scale) &     --1.0         &  \ref{eqn:gasabundance}         \\ 	
 	$[X_\mathrm{Mg}/\mathrm{H}]$                              &Mg dust depletion factor (log scale) &   --0.5           &   \ref{eqn:gasabundance}          \\
 	$D$                                & Spatial extent of the line-absorbing gas in winds                   &   5 kpc           &   \ref{eqn:shellmodelofwindmass}   \\
 	 \shortstack{$A_\mathrm{wind}$ \\ \ }        & \shortstack{Surface area covered by wind   \\ \ }         &   \shortstack{ \\ $= 2\times\,$Surface area$\times$\\ \ \ Covering fraction}   &  \shortstack{\ref{eqn:shellmodelofwindmass} \\ \ }  \\ \hline
 	\end{tabular}
 \end{table*}
 
\section{Winds \& Star Formation Rates}
\label{sec:discussphysicalimpacts}

The SFR density map in Figure~\ref{fig:sfrdensity} shows that both the Central and Arc regions have SFR densities of around 0.2 $M_\sun\,\mathrm{yr}^{-1}\, \mathrm{kpc}^{-2}$ (Table \ref{tab:targetprop}).  Given that the winds likely originate from both regions (\S\ref{sec:discusswindorigin}), we speculate that the winds are driven by the spatially extended star formation which covers both the Central and Arc regions of the galaxy.

We note that the SFR densities of both regions are above 0.1 $M_\sun\,\mathrm{yr}^{-1}\, \mathrm{kpc}^{-2}$, which is the threshold for local starbursts to launch strong winds (\citealt{Heckman2002,Heckman2015}). Winds are indeed detected in both regions for our study, consistent with this threshold measured from low redshifts. 
 
 \begin{table*}
	\renewcommand{\thetable}{\arabic{table}}
	\caption{The hydrogen column densities, column-density-weighed velocities, mass outflow rates, and mass loading factors}
	\label{tab:massratesfr}
	\centering
	\begin{tabular}{lll}
		\hline
		\hline
		Quantity & Central region & Arc region  \\
		\hline 
		Hydrogen Column Density ($\log\,N_\mathrm{H}$)  &  & \\
		\hspace{0.2in} estimated from \ion{Fe}{2} &          $20.2^{+0.2}_{-0.2}\ [\mathrm{cm}^{-2}]$                 &         $20.1^{+0.1}_{-0.2}\  [\mathrm{cm}^{-2}]$                     \\
		\hspace{0.2in} estimated from \ion{Mg}{2}$^a$ &          $19.3^{+0.3}_{-0.2}\ [\mathrm{cm}^{-2}]$                 &                  $>19.5\ [\mathrm{cm}^{-2}]$                   \\	Column-density-weighted velocity inferred from Fe II ($\bar{v}_\mathrm{wind}$) & $266^{+14}_{-14}$ km$\cdot$s$^{-1}$ & $228^{+7}_{-6}$ km$\cdot$s$^{-1}$ \\
		Mass outflow rate estimated from \ion{Fe}{2}$^{b}$ &  $4^{+4}_{-2}\ [M_\sun\,\mathrm{yr}^{-1}]$     &  $3^{+2}_{-1}\ [M_\sun\,\mathrm{yr}^{-1}]$    \\ 
		Mass loading factor estimated from \ion{Fe}{2}$^{b}$ & $0.2^{+0.6}_{-0.1}$ & $0.2^{+0.4}_{-0.1}$ \\
		\hline
	\end{tabular}
	\begin{flushleft}
	
		\tablenotetext{a}{Likely underestimated due to line saturation (Appendix \ref{sec:mgfediscrepancy})}
		\tablenotetext{b}{The uncertainties quoted for the mass outflow rates and mass loading factors do not include the systematic uncertainties due to the assumptions made for their calculations. These systematic uncertainties may be an order of magnitude or more.}		
		
	\end{flushleft}
	
\end{table*}

\section{Estimated Mass Outflow Rates \& Mass Loading Factors}

\label{sec:massoutflowratemassloading}

We estimate the mass outflow rates and mass loading factors for the two spatial regions. The estimated mass outflow rates are $4\,M_\sun\,\mathrm{yr}^{-1}$   and  $3\,M_\sun\,\mathrm{yr}^{-1}$ for the Central and Arc regions, respectively, and the corresponding mass loading factors are both around 0.2. However, we caution that these values are only accurate to approximately an order of magnitude.

\subsection{Hydrogen Column Densities}

\label{sec:hcolumndensity}


Hydrogen column densities ($N_\mathrm{H}$) are a prerequisite for the calculation of mass outflow rates. They are inferred from the \ion{Fe}{2} or \ion{Mg}{2} column densities, $N_{\rm Fe\,II}$ and $N_{\rm Mg\,II}$.  The calculation of $N_\mathrm{H}$ from $N_\mathrm{Fe\,II}$ is detailed below, whereas that from $N_\mathrm{Mg\,II}$ follows the same steps.

Two quantities are needed to infer $N_\mathrm{H}$ from $N_\mathrm{Fe\,II}$, namely the gas-phase elemental abundance of Fe ($N_\mathrm{Fe,\,gas}/N_\mathrm{H}$), and the ionization fraction of Fe in the gas phase ($N_\mathrm{Fe\,II}/N_\mathrm{Fe,\,gas}$):  
\begin{equation}
	N_\mathrm{H} = \frac{N_\mathrm{Fe\,II}}{ (N_\mathrm{Fe\,II}/N_\mathrm{Fe,\,gas})\cdot (N_\mathrm{Fe,\,gas}/N_\mathrm{H})}.
	\label{eqn:ionizationcorrection}
\end{equation}
The gas-phase elemental abundance is determined from the total abundance of Fe in gas and dust, $\log\,$(Fe/H)$_\mathrm{total}$, and the dust depletion factor, $[X_\mathrm{Fe}/\mathrm{H}]$:

\begin{align}
	\log (N_\mathrm{Fe,\,gas}/N_\mathrm{H}) & = \log (\mathrm{Fe/H})_\mathrm{total} - [X_\mathrm{Fe}/\mathrm{H}]. \label{eqn:gasabundance}
\end{align}
We assume the ionization fraction values to be 1.0, the total metal abundance to be solar (\citealt{Asplund2009,Chisholm2016}), and the dust depletion factors to be the average values for the Milky Way ISM ((\citealt{Jenkins2009}). Their numerical values are listed in  Table~\ref{tab:assumptions}. 



Following the calculations outlined above, we are able to infer two values of $N_\mathrm{H}$ from \ion{Fe}{2} and \ion{Mg}{2}, respectively, for each of the Central and Arc regions.  These values are listed in Table \ref{tab:massratesfr}. However, we caution that the $N_\mathrm{H}$ values inferred from \ion{Mg}{2} are likely substantially underestimated due to the assumption about the \ion{Mg}{2} ionization fraction and/or the density inhomogeneity of the outflowing gas, which we explain in detail in Appendix \ref{sec:mgfediscrepancy}. As a result, \emph{we only adopt the $N_\mathrm{H}$ values inferred from \ion{Fe}{2} for the rest of this paper}. 

\subsection{Mass Outflow Rates \& Mass Loading Factors}

To estimate the mass outflow rates from each region of the galaxy, we follow the same steps as in \cite{Rubin2014}. They assume that the winds traced by \ion{Mg}{2} or \ion{Fe}{2} are in the form of a continuous flow from their launching sites to a radial distance $D$, and they have an average radial velocity $\bar{v}_\mathrm{wind}$. The equation for the mass outflow rate is given in \S 8.4.2 of \cite{Rubin2014}:
\begin{equation}
	\dot{M}_\mathrm{wind} = 1\, M_\sun\ \mathrm{yr}^{-1} 
	\frac{N_\mathrm{H}}{10^{20}\,\mathrm{cm^{-2}}}
	\frac{A_\mathrm{wind}}{45\, \mathrm{kpc}^2} \frac{\bar{v}_\mathrm{wind}}{300\,\mathrm{km\cdot s^{-1}}} \frac{\mathrm{5\,kpc}}{D},
\label{eqn:shellmodelofwindmass}
\end{equation}
where $N_\mathrm{H}$ is the hydrogen column density  inferred from the previous subsection and $A_\mathrm{wind}$ is the surface area of the galaxy covered by the wind. The area $A_\mathrm{wind}$ equals the geometric surface area of each spatial region listed in Table \ref{tab:targetprop} multiplied by the gas covering fraction $C_f$ from Table \ref{tab:lineprofilefitresult} and an additional factor of 2 (\citealt{Rubin2014}). The term $\bar{v}_\mathrm{wind}$ is estimated as follows:
\begin{equation}
\bar{v}_\mathrm{wind} = \frac{\int_{-450\ \mathrm{km}\cdot \mathrm{s}^{-1}}^{-150\ \mathrm{km}\cdot \mathrm{s}^{-1}} v\cdot  N_{v\mathrm{,\,ion}}(v)\ d\,v}{\int_{-450\ \mathrm{km}\cdot \mathrm{s}^{-1}}^{-150\ \mathrm{km}\cdot \mathrm{s}^{-1}}  N_{v\mathrm{,\,ion}}(v)\ d\, v},
\label{eqn:weightedwindvelocity}
\end{equation}
where $N_{v\mathrm{,\,ion}}(v)$ is the \ion{Fe}{2} column density as a function of velocity shown in Figure \ref{fig:nionprof}. Values of $\bar{v}_\mathrm{wind}$ for the two regions are listed in Table \ref{tab:massratesfr}. The term $D$ is assumed to be 5 kpc following \cite{Rubin2014}, although this number might underestimate the truth by factor of three or four according to recent observations of galactic winds at $z\sim1$  (\citealt{Finley2017,Burchett2020,Rupke2019,Burchett2020,Zabl2021}; but see also \citealt{Erb2012,Tang2014} which favor $D \simeq 5$ kpc).


The inferred mass outflow rate is $4^{+4}_{-2}\ M_\sun\,\mathrm{yr}^{-1}$ for the Central region and $3^{+2}_{-1}\ M_\sun\,\mathrm{yr}^{-1}$ for the Arc region.  We caution that the quoted uncertainties do not include systematic uncertainties, and that the mass outflow rates calculated here only serve as \emph{a rough  estimate} of their true values. The full uncertainties of the mass outflow rates are likely substantial, one order of magnitude or more, due to the systematic errors in the gas metallicity ($\pm 0.4$ dex; \citealt{Chisholm2016, Chisholm2018}), ionization fraction ($\pm 1.0$ dex; \citealt{Murray2007,Narayanan2008,Giavalisco2011,Rubin2014,Crighton2015}), dust depletion ($\pm 0.2$ dex; \citealt{DeCia2016,Jones2018,Wendt2020}), and spatial extent of the winds ($\pm 0.5$ dex; \citealt{Rupke2019,Burchett2020,Zabl2021}).

Finally, the mass loading factors are defined as the ratio between the mass outflow rate and the SFR of each spatial region. They are estimated to be 0.2 for both regions.  However, due to the uncertainties of the mass outflow rates, the uncertainties of the mass loading factors can again be one order of magnitude or more.  

\section{Comparison With Previous Studies \& Future Observations with \emph{JWST}}

\label{sec: compare_with_literature}

The main result of this paper, i.e. that winds are launched from the entire spatial extent of the galaxy, is broadly consistent with three other similar studies at $z\gtrsim 1$.  \cite{Bordoloi2016} study a gravitationally lensed star-forming galaxy with four bright star-forming clumps at $z=1.7$ and find cool outflows from all the four with comparable gas column densities and mass outflow rates.  Two other studies by \cite{James2018} and \cite{RickardsVaught2019} measure the equivalent widths of the low-ionization absorption lines tracing the cool outflows along multiple sightlines toward a star-forming galaxy at $z=2.4$ and a star-forming galaxy at $z=0.7$, respectively. They find the line equivalent widths to be comparable among different sightlines, indicating winds from the entire spatial extents of the galaxies. In addition, the three studies also suggest that the spatially extended star formation inside the massive star-forming galaxies at this cosmic epoch drives the observed winds from the entire faces of the galaxies, which agrees qualitatively with this work and several recent  hydrodynamic simulations (c.f. fig.~15 of \citealt{Grand2019} and fig.~12 of \citealt{Nelson2019b}). 

Notwithstanding the overall consistent findings of the studies mentioned above, more observations are needed to further our understanding of the roles of galactic winds in galaxy formation at $z\gtrsim 1$. Here we identify two specific aspects to be explored in the future. First, the connection between star formation and the launching of galactic winds remains to be studied at spatially scales smaller than those by current studies.  Current observations of the cool galactic winds at $z \gtrsim 1$ only reach spatial resolutions of around 6 kpc (0.8\arcsec) due to seeing. To reach around 1 kpc, for example, a spatial resolution of 0.2\arcsec\ will be needed. Second, the impacts of winds on the formation of the morphological structures of galaxies remain to be understood. The cosmic epoch of $z\gtrsim 1$ sees the emergence of these morphological structures, such as bulges and disks (e.g., \citealt{Kassin2007,Kassin2012,Wuyts2011b,Patel2013,Conselice2014,Huertas-Company2016, Simons2017, Costantin2022}).  To understand the impacts of winds on these structures, we will need a large sample of galaxies with a broad range of morphology and measure the winds from their disks, bulges, etc. 

Such high spatial resolution and multi-object observations are soon to be made possible by the \emph{JWST} NIRSpec instrument. The Mirco-Shutter Assembly onboard the instrument will enable deep spatially resolved spectroscopic observations of more than 30 galaxies at $z\gtrsim 1$ simultaneously and deliver 2D maps of several spectral lines tracing winds, including \ion{Fe}{2}, \ion{Mg}{2}, and Na I D, at a spatial resolution of around 0.2\arcsec.


\section{Conclusion}

\label{sec:conclusion}

We study the extended winds from a massive star-forming galaxy at $z=1.3$ using deep spectra from the DEIMOS spectrograph on Keck. The morphology of the galaxy is indicative of a recent merger.  Gas kinematics indicate a dynamically complex system with velocity gradients ranging 0--60 $\mathrm{km}\cdot\mathrm{s}^{-1}$. For this galaxy, we measure the properties of the cool outflowing gas ($\sim10^4$ K) from \ion{Fe}{2} and \ion{Mg}{2} absorption lines.  This is done for two regions of the galaxy: a dust-obscured center (``Central region"), and an extended arc (``Arc region") which is around 7 kpc away from the center. These regions make the galaxy visually resemble a ``Baltimore Oriole's Nest," which is the nickname we give it.  Our main results are as follows:

\begin{itemize}
	
	\item  Outflows are detected in both regions of the galaxy according to the observed blueshifted \ion{Fe}{2} and \ion{Mg}{2} absorption lines. For both regions, the wind velocities are in the range of 100--450 km$\cdot$s$^{-1}$ (\S \ref{sec:lineprofqualitativediscuss})
	
	\item The \ion{Fe}{2} absorption line profiles of the two regions have similar shapes. The inferred \ion{Fe}{2} column densities of the outflowing gas for the two regions are also similar: $10^{14.7^{+0.2}_{-0.2}}\, \mathrm{cm}^{-2}$  for the Central region and $10^{14.6^{+0.1}_{-0.2}}\, \mathrm{cm}^{-2}$ for the Arc region (\S  \ref{sec:ewmeasurement} \&  \ref{sec:functionfittinglines}).
	
	\item Our results prefer a scenario in which galactic winds are launched from both regions of the galaxy (\S \ref{sec:discusswindorigin}). 
	
	\item The mass outflow rates are estimated to be $4\,M_\sun\,\mathrm{yr}^{-1}$ and  $3\,M_\sun\,\mathrm{yr}^{-1}$ for the Central and Arc regions, respectively (\S \ref{sec:massoutflowratemassloading}). However, the systematic uncertainties of these values may be one order of magnitude or more.
	
\end{itemize}

We speculate that winds are most likely driven by the spatially extended star formation of the galaxy (\S \ref{sec:discussphysicalimpacts}). The SFR densities are similar in the two regions, both around  0.2$\,M_\sun\,\mathrm{yr}^{-1}\, \mathrm{kpc}^{-2}$, according to the SED modeling.

This work is the pilot study of a larger program, in which we will use deep DEIMOS data to study spatially resolved properties of the winds from a large sample of massive star-forming galaxies at $z\sim 1$. Massive star-forming galaxies at this cosmic epoch are expected to have extended star formation similar to the galaxy studied in this work (e.g., \citealt{Wang2017a,Liu2018,Tacchella2018, Morselli2019,Nelson2019a}), and therefore are expected to have winds launched from extended areas. This work is also a path finder for future studies with the \emph{James Webb Space Telescope}, with which spatially resolved spectroscopic observations of winds can be conducted with high spatial resolution and spectral sensitivity at $z\gtrsim 1$.

\acknowledgments

WW and SAK would like to acknowledge support from NASA's Astrophysics Data Analysis Program (ADAP) grant number 80NSSC20K0760 and an RSAC grant from the Space Telescope Science Institute. DCK acknowledges support from the National Science Foundation grant AST-1615730. ECC and PG acknowledge support from the National Science Foundation grant AST-1616540.  ECC is supported by a Flatiron Research Fellowship at the Flatiron Institute. The Flatiron Institute is supported by the Simons Foundation. CP is supported by the Canadian Space Agency under a contract with NRC Herzberg Astronomy and Astrophysics. HY acknowledges support from the Research Fund for International Young Scientists of NSFC (11950410492).

Some of the data presented herein were obtained at the W.~M.~Keck Observatory, which is operated as a scientific partnership among the California Institute of Technology, the University of California and the National Aeronautics and Space Administration. The Observatory was made possible by the generous financial support of the W.~M.~Keck Foundation. The authors wish to recognize and acknowledge the very significant cultural role and reverence that the summit of Maunakea has always had within the indigenous Hawaiian community.  We are most fortunate to have the opportunity to conduct observations from this mountain. This work is based on observations taken by the CANDELS Multi-Cycle Treasury Program with the NASA/ESA HST, which is operated by the Association of Universities for Research in Astronomy, Inc., under NASA contract NAS5-26555.

We are grateful for Mr.~John Anes offering his photo of the Baltimore Oriole, which is adopted in this paper. We wish his children many beautiful star-filled nights with their new telescope. The authors would also like to thank Camilla Pacifici for her keen eye in relating the galaxy in this paper to a bird in a nest. We thank Sylvain Veilleux, J.~X.~Prochaska and John Chisholm for the inspirational discussions at various stages of this project. WW also wants to thank Guido Roberts-Borsani for providing a table of mass loading factor values from his paper and Robert Grand for providing a snapshot of the Auriga galaxy simulation. 

This research made use of Astropy,\footnote{http://www.astropy.org} a community-developed core Python package for Astronomy \citep{astropy2013,astropy2018}.


\clearpage


\appendix

\begin{figure}
	\centering
	\includegraphics[width = 6.8 in]{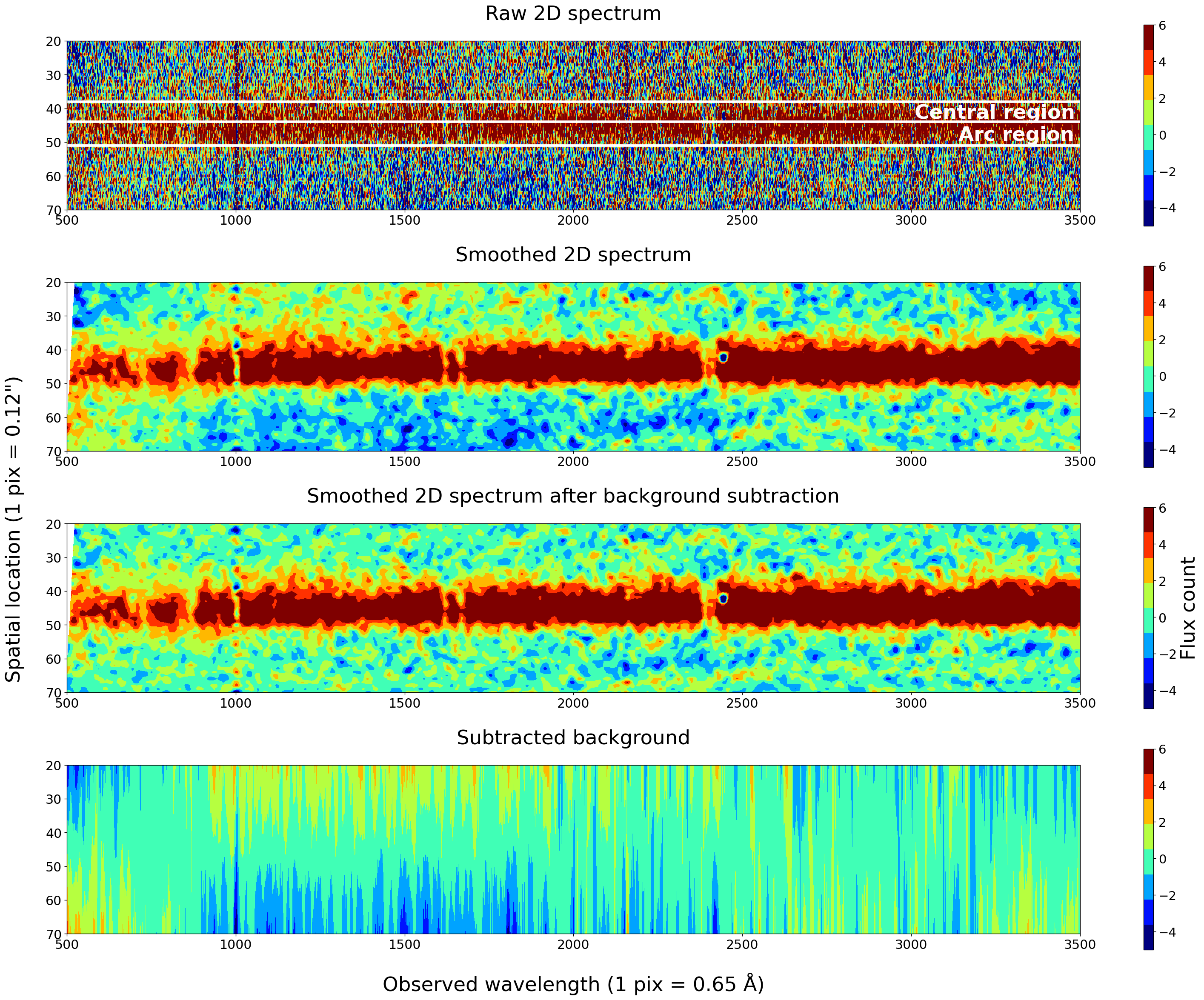}
	\caption{ \emph{First panel:} The 2D spectrum obtained with a slit position angle of 10$\degr$. The spatial direction is vertical and the wavelength direction is horizontal in the figure, and the units of the two axes are pixel numbers. The galaxy is located in rows 35--55. The ranges where the 1D spectra of the Central and Arc regions are extracted are indicated in the figure. Color bars indicate the flux of each pixel, in units of flux counts per hour. \emph{Second panel:}  The same 2D spectrum but smoothed along the spatial and spectral directions.  A significant spatial variation (vertical direction in the figure) in the background flux can be easily seen from this smoothed version of the spectrum. This is especially the case for columns 1000--2000, where the flux varies from around -3 to around 3 from the bottom to the top of the spectrum.  \emph{Third panel:} The 2D spectrum after the background is subtracted.  \emph{Fourth panel:} The background that is subtracted out. This correction brings a flux change of no more than 10\% to the resulting 1D spectra relative to the continuum. Details about modeling the spatially varying background can be found in Appendix \ref{appendix:backgroundissue}. 
		\label{fig:skysubtraction} }
\end{figure}

\vspace{-0.8cm}

\section{Spectral Background Subtraction}
\label{appendix:backgroundissue}

The background in the DEIMOS 2D spectra  varies along the slit. This spatial variation is not taken into account by the standard data reduction pipeline. We present a 2D spectrum in Figure \ref{fig:skysubtraction} to show this variation. The spatial direction is vertical and the wavelength direction is horizontal. The galaxy studied in this paper is located in rows 35--55. As can be seen from the second panel in the figure where the 2D spectrum is smoothed to show variations on larger scales, the surrounding background varies significantly along the spatial direction. This is especially the case for columns 1000--2000, where the flux varies from around -3 to 3.  The spatial variation is likely due to light leaking from neighboring slits on the same DEIMOS mask and/or scattered from other astronomical or artificial sources during the observations. 

To remove spatial variation, we assume the background flux is a linear function of the row number. The fit is done for every column of the 2D spectrum. This results in a 2D model for the background. The 2D model is then smoothed with a Gaussian of RMS of 10 pixels (6.5 \AA) along the wavelength direction. The smoothed model is then subtracted from the original 2D spectrum, and the result is shown in the middle panel of Figure \ref{fig:skysubtraction}. It has a more uniform background along the spatial direction than before the subtraction. We find that this background removal brings a flux change of no more than 10\% to the resulting 1D spectra relative to the spectral continuum for both spatial regions in the galaxy we study.

\section{Beam Smearing}
\label{appendix:beamsmearing}
 
\begin{figure*}
 	\centering
 	\includegraphics[width = 6.3 in]{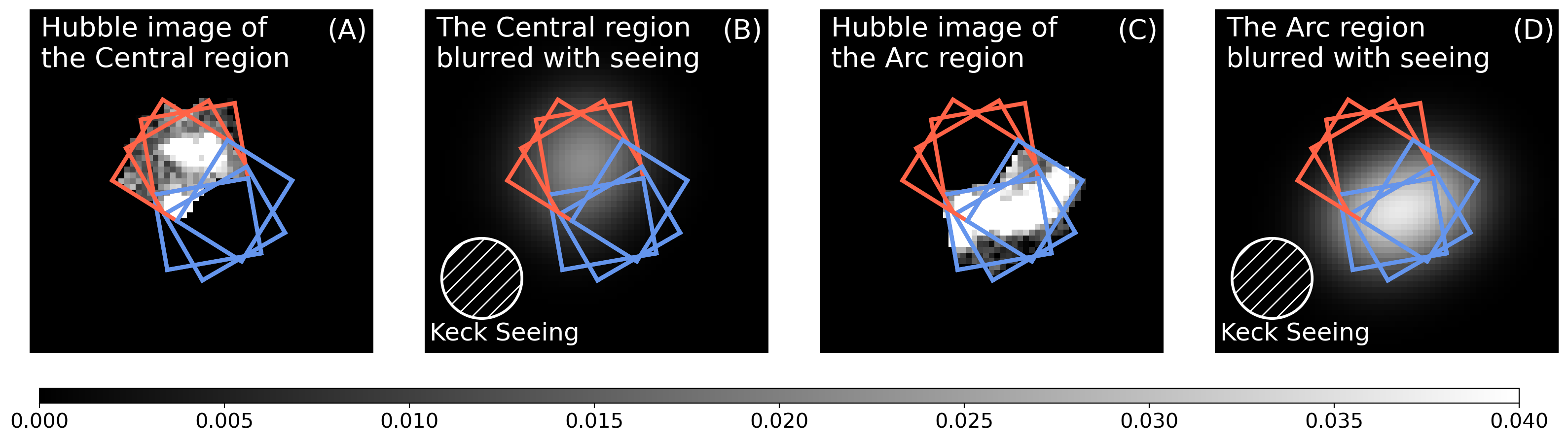}
 	\caption{\emph{HST} images of the Central and Arc regions in the ACS/F606W band are shown in Panels A and C, respectively.  They are blurred by the ground-based seeing in Panels B and D, respectively. Red and blue rectangles indicate the bounds of the two regions used in our analysis. The seeing has an FWHM of 0.86\arcsec, which is indicated by the hatched circles. On the blurred images, 70\% (80\%) of the light collected in the Central (Arc) region originates in the high-resolution image of the same region. This means that the beaming smearing effect is not significant and the two regions can be resolved in the spectral observations. The color bar indicates the image surface brightness in arbitrary units.
 		\label{fig:beamsmearingissue}}
 \end{figure*}

We show the influence of atmospheric seeing during the observations, and validate that the galaxy can be sufficiently resolved into two regions, the Central region and the Arc region.  To do this, we convolve a high-resolution \emph{HST} ACS/F606W image of the galaxy with the ground-based seeing (FWHM=0.86\arcsec).  In Figure \ref{fig:beamsmearingissue}, we show the original resolution \emph{HST} images of just the Central and Arc regions in Panels A and C, respectively.  The convolved images of these regions are shown in Panels B and D, respectively. About 70\% of the light from the Central region originates in the high-resolution image of the same region.  This value reaches 80\% for the Arc region.  Therefore, the two spatial regions are indeed resolved in the Keck observations.

 \section{Measuring redshifts and searching for outflows from the [\ion{O}{2}] emission lines }
 
 \label{appendix:oiimeasurement}
 
\begin{figure}
 	\centering
 	\includegraphics[width = 6.3 in]{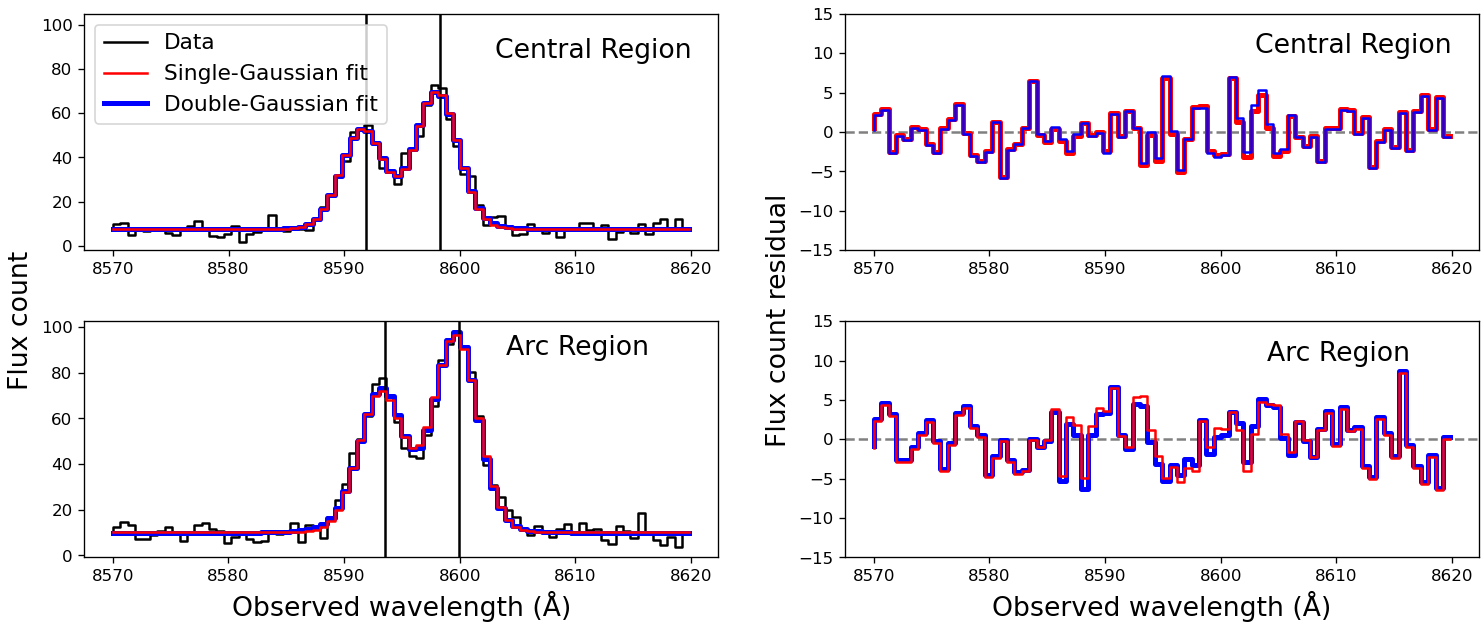}
 	\caption{We fit the observed line profiles of the [\ion{O}{2}] $\lambda\lambda$ 3727/3729\,\AA\ doublet with Gaussian functions to infer the redshifts and look for outflows in the warm ionized phase for the Central and the Arc regions.  Left panels show the observed emission line profiles (black) along with the models (red and blue), and right panels show residuals of the fitting.  For both regions, each line in the doublet can be well fit with just one Gaussian (red lines). Centroids of the best-fit Gaussians are used to infer the redshift of each spatial region. Models that invoke two Gaussian components for each line are also presented (blue lines) and they do not improve the fitting compared to the single-Gaussian models.  This means that the broad wings which are signatures of the outflowing gas are not present in the observed [\ion{O}{2}] line profiles, and the [\ion{O}{2}] lines only trace the ISM. All the Gaussian models in the figure have been convolved with the DEIMOS line spread function to account for the instrument resolution.   \label{fig:o2fitting} }
 \end{figure}

 The redshifts of the Central and Arc regions are measured from the [\ion{O}{2}] $\lambda\lambda$ 3727/3729\,\AA\ nebular emission lines. For each region, the measurement is done by fitting each of the two [\ion{O}{2}] lines with Gaussian profiles. The observed line profiles (black curves) and the Gaussian fits (red and blue curves) are presented in Figure \ref{fig:o2fitting}.
 
Before we determine the redshift values, we inspect the fitting results and examine whether the [\ion{O}{2}] emission lines are only from the galaxy ISM or they are also from the ionized gas outflows.  If the outflows are present, broad wings would be expected in the line profiles and each of the [\ion{O}{2}] line doublet would be better fit with two Gaussians (double-Gaussian model) than a single Gaussian (single-Gaussian model) (e.g., \citealt{Genzel2011, Newman2012, ForsterSchreiber2014,ForsterSchreiber2019, Zakamska2014, Davies2019}). We show in the left panels of Figure \ref{fig:o2fitting} that the double-Gaussian fits are indistinguishable from the single-Gaussian fits. In addition, as shown on the right, the residuals for single-Gaussian fits do not have any features. Therefore, single-Gaussian fits are sufficient  and no signature of outflows are present in the [\ion{O}{2}] line profiles.
 
 After concluding that the [\ion{O}{2}] lines only trace the ISM of the galaxy, we use the single-Gaussian fits to determine the observed wavelengths of the line centroids and infer the systemic redshifts. The inferred redshifts are 1.3060 and 1.3063 for the Central and Arc regions, respectively. The velocity dispersion values of the  [\ion{O}{2}] lines shown in Figure \ref{fig:o2fitting} are measured to be $65\,\mathrm{km}\,\mathrm{s}^{-1}$  and $67\,\mathrm{km}\,\mathrm{s}^{-1}$ for the Central and Arc regions, respectively, before the instrument line spread function is subtracted, and $40\,\mathrm{km}\,\mathrm{s}^{-1}$  and $42\,\mathrm{km}\,\mathrm{s}^{-1}$  afterwards, correspondingly.
 
 \section{Relation Between Line Equivalent Width and Column Density}
 
 \label{sec:ewversusn_physics}
 
We provide a brief explanation on the relation between the equivalent widths of the absorption lines and the ion column density. We explain that the column densities can be constrained from the relative difference in the equivalent widths of the two absorption lines that have different oscillator strengths.
 
A larger (smaller) difference in the equivalent widths of the two lines indicates a lower (higher) ion column density. This is because the two lines reach the optically thick regime (saturation) at different column densities. An intuitive example is shown in the middle and bottom panels of Figure \ref{fig:lineprofiledemo}. The \ion{Mg}{2} $\lambda\lambda$ 2796/2803 \AA\ doublet is shown as the example, and arguments for the \ion{Fe}{2} doublets are similar. The two \ion{Mg}{2} lines have different oscillator strengths: $f_{2796}=0.61$ and $f_{2803}=0.31$. Because the optical depth is proportional to the column density times their $\lambda f$ values (Equation \ref{eqn:tauversusfN}), their optical depths differ by a constant factor of $(\lambda_{2803} f_{2803})/(\lambda_{2796} f_{2796})=0.5$.  At a low column density, the \ion{Mg}{2} 2803 \AA\ is optically thin and weaker than \ion{Mg}{2} 2796 \AA\ (middle left and bottom left of Figure \ref{fig:lineprofiledemo}). Therefore, the equivalent width of \ion{Mg}{2} 2803 \AA\  is smaller than that of \ion{Mg}{2} 2796 \AA: $\mathrm{EW}_{2803}<\mathrm{EW}_{2796}$. The relation of equivalent widths changes when the \ion{Mg}{2} column density increases to a value (around $10^{14.5}\mathrm{cm}^{-2}$ for this work) such that both lines become optically thick ($\tau_{2796}\geq 5$, $\tau_{2803}\geq 5$). This case is shown on the middle right and bottom right of Figure \ref{fig:lineprofiledemo}. The two \ion{Mg}{2} absorption lines fully saturate and reach similar equivalent widths: $\mathrm{EW}_{2803} \simeq \mathrm{EW}_{2796}$. Changing the gas covering fraction does not change the arguments made above. More detailed discussion can be found in textbooks like \cite{Draine2011}.

 \section{Discrepancies Between the Hydrogen Column Densities Inferred from \ion{F\lowercase{e}}{2} and \ion{M\lowercase{g}}{2}}
 
 \label{sec:mgfediscrepancy}
We argue here that the hydrogen column densities inferred from \ion{Fe}{2} in the main text are better estimates than those inferred from \ion{Mg}{2}.  In \S \ref{sec:hcolumndensity}, we find that the hydrogen column density inferred from Fe is around a factor of 8 or 0.9 dex higher than that inferred from Mg for the Central region. Similar discrepancies are also found in the literature (e.g., \citealt{Rigby2002,Churchill2003,Narayanan2008,Rubin2010,Bordoloi2016}). Two possible explanations exist for the discrepancy.
 
First, the discrepancy is likely caused by uncertainties in the conversion from the ion column density to the hydrogen column density. The conversion is dependent on the Fe and Mg dust depletion factors and the ionization fractions. The assumed values of the dust depletion factors are not likely to be the culprit of the discrepancy, however. This is because the depletion factors measured from observations have intrinsic scatters of no more than 0.15 dex (\citealt{Jenkins2009}; Fig.~9 of \citealt{DeCia2016}), which result in an uncertainty of no more than 0.15 dex in the inferred hydrogen column densities, significantly smaller than the 0.9 dex discrepancy. It is possible to resolve the discrepancy by adjusting the assumed \ion{Mg}{2} ionization fraction value from 1.0 (which is adopted in the main text) to 0.1--0.2, while keeping the \ion{Fe}{2} ionization fraction value adopted in the main text unchanged. Such adjustment, although yet to be justified by physical ionization models, would indicate that the hydrogen column densities inferred from \ion{Fe}{2} are closer to truth, whereas those originally inferred from \ion{Mg}{2} in the main text are significantly underestimated.
 
The second explanation is that the hydrogen column density inferred from \ion{Mg}{2} is underestimated due to a population of dense clouds or radial filaments in the winds that is optically thick for the \ion{Mg}{2} lines. Such clouds are optically thick for the \ion{Mg}{2} lines because the  oscillator strengths of the lines are high. For example, the \ion{Mg}{2} 2803\,\AA\ line has a 5 and 12 times higher oscillator strength than the \ion{Fe}{2} 2896\,\AA\ and \ion{Fe}{2} 2374\,\AA\ lines, respectively (Table \ref{tab:lineflambda}). As a result, the \ion{Mg}{2} lines saturate at a lower column density than the \ion{Fe}{2} lines and the observed \ion{Mg}{2} line profiles are dominated by sightlines that probe the low column density gas, not the dense clouds. The hydrogen column density inferred from \ion{Mg}{2} will be underestimated because the dense clouds are missed. This effect for \ion{Mg}{2} is reminiscent of the ``hidden saturation'' effect discovered for the ISM of local galaxies: \cite{James2014} find that the hydrogen column densities inferred from absorption lines with higher oscillator strengths are lower than those inferred from lines with lower oscillator strengths. They have an explanation similar to ours, i.e., that the discrepancy is caused by density inhomogeneities in the gas. Same as the first explanation, this second explanation indicates that the hydrogen column densities inferred from \ion{Mg}{2} are substantially underestimated.
 
 In summary, both explanations for the discrepancy indicate that the hydrogen column densities inferred from \ion{Fe}{2} in our calculation are closer to truth, whereas those inferred from \ion{Mg}{2} are substantially underestimated. \emph{Therefore, only the hydrogen column densities inferred from \ion{Fe}{2} are adopted in this paper, for both regions of the galaxy}.

\bibliography{wind}{}
\bibliographystyle{aasjournal}



 \end{document}